\documentclass[10pt,journal,compsoc]{IEEEtran}
\ifCLASSOPTIONcompsoc
  \usepackage[nocompress]{cite}
\else
  \usepackage{cite}
\fi
\ifCLASSINFOpdf
   \usepackage[pdftex]{graphicx}
\else
   \usepackage[dvips]{graphicx}
\fi

\ifCLASSOPTIONcompsoc
 \usepackage[font=footnotesize,labelfont=sf,textfont=sf]{subfig}
\else
 \usepackage[font=footnotesize]{subfig}
\fi

\usepackage{amsmath,amssymb,amsfonts}
\usepackage{acronym}
\usepackage{algorithmic}
\usepackage{array}
\usepackage{url}
\newcommand\MYhyperrefoptions{bookmarks=true,bookmarksnumbered=true,
pdfpagemode={UseOutlines},plainpages=false,pdfpagelabels=true,
colorlinks=true,linkcolor={black},citecolor={black},urlcolor={black},
pdftitle={Leaking Secrets through Modern Branch Predictors in the Speculative World},%
pdfsubject={Branch predictor speculative side channel},%
pdfauthor={Md Hafizul Islam Chowdhuryy},%
pdfkeywords={Branch Predictor, Transient Execution Attack, Side Channel, Speculative Execution}}%
\ifCLASSINFOpdf
\usepackage[\MYhyperrefoptions,pdftex,breaklinks]{hyperref}
\else
\usepackage[\MYhyperrefoptions,breaklinks=true,dvips]{hyperref}
\usepackage{breakurl}
\fi

\usepackage{textcomp}
\usepackage{xcolor}
\usepackage{xspace}
\usepackage{listings,multicol}
\usepackage{enumitem}
\usepackage{multirow}
\usepackage{tabularx}
\usepackage{soul}
\usepackage{pifont}
\usepackage[linesnumbered,ruled,vlined]{algorithm2e}
\usepackage{booktabs}
\usepackage[utf8]{inputenc}

\usepackage{breakurl}

\usepackage[T1]{fontenc}
\usepackage{listings}
\usepackage{color}
\makeatletter
\newcommand\BeraMonottfamily{%
  \def\fvm@Scale{0.80}%
  \fontfamily{fvm}\selectfont%
}
\makeatother

\definecolor{darkviolet}{RGB}{157,0,129}
\definecolor{darkgreen}{RGB}{31,86,0} 
\definecolor{darkblue}{RGB}{0,104,133}
\definecolor{darkgrey}{rgb}{0.5,0.5,0.5}
\definecolor{lightblue}{rgb}{0.4,0.4,1}
\definecolor{stringColor}{rgb}{0.16,0.00,1.00}
\definecolor{fieldColor}{rgb}{0.16,0.00,1.00}
\definecolor{annotationColor}{rgb}{0.39,0.39,0.39}
\definecolor{keywordColor}{rgb}{0.50,0.00,0.33}
\definecolor{commentColor}{rgb}{0.25,0.50,0.37}
\definecolor{javadocColor}{rgb}{0.25,0.37,0.75}
\definecolor{jTagColor}{rgb}{0.50,0.62,0.75}
\definecolor{eTagColor}{rgb}{0.50,0.62,0.75}
\definecolor{lineNumberColor}{rgb}{0.47,0.47,0.47}

\SetCommentSty{mycommfont}

\newcommand{\mytitle}{BranchSpectre\xspace}

\begin{document}
\title{{Leaking Secrets through Modern Branch Predictors in the Speculative World}}

\author{Md~Hafizul~Islam~Chowdhuryy, ~\IEEEmembership{Student Member,~IEEE,} and ~Fan~Yao, ~\IEEEmembership{Member,~IEEE}%
\IEEEcompsocitemizethanks{\IEEEcompsocthanksitem M. Chowdhuryy\IEEEauthorrefmark{2} and F. Yao\IEEEauthorrefmark{4} are with the Department
of Electrical and Computer Engineering, University of Central Florida, Orlando,
FL.\protect\\
E-mail: \IEEEauthorrefmark{2}reyad@knights.ucf.edu, \IEEEauthorrefmark{4}fan.yao@ucf.edu
}

}

\bstctlcite{IEEEexample:BSTcontrol}

\IEEEtitleabstractindextext{%
\begin{abstract}

Transient execution attacks that exploit speculation have raised significant concerns in computer systems. 
Typically, branch predictors are leveraged to trigger mis-speculation in transient execution attacks. 
In this work, we demonstrate a new class of speculation-based attacks that targets the branch prediction unit (BPU). We find that speculative resolution of conditional branches (i.e., in nested speculation) alter the states of pattern history table (PHT) in modern processors, which are not restored after the corresponding branches are later squashed. Such characteristic allows attackers to exploit the BPU as the secret transmitting medium in transient execution attacks. To evaluate the discovered vulnerability, we build a novel attack framework, \emph{\mytitle}, that enables exfiltration of unintended secrets through observing speculative PHT updates (in the form of covert and side channels). We further investigate the PHT collision mechanism in the history-based predictor and the branch prediction mode transitions in Intel processors. Built upon such knowledge, we implement an ultra-high speed covert channel (\emph{\mytitle-cc}) as well as two side channels (i.e., \emph{\mytitle-v1} and \emph{\mytitle-v2}) that merely rely on BPU for mis-speculation trigger and secret inference in the speculative domain. Notably, \mytitle side channels can take advantage of much simpler code patterns than those used in Spectre attacks. We present an extensive \mytitle code gadget analysis on a set of popular real-world application code bases followed by a demonstration of side channel attack on OpenSSL. The evaluation results show substantially wider existence and higher exploitability of \mytitle code patterns in real-world software. 
Finally, we discuss several secure branch prediction mechanisms that can mitigate transient execution attacks exploiting modern branch predictors.

\end{abstract}

}

\maketitle

\IEEEdisplaynontitleabstractindextext

\IEEEpeerreviewmaketitle

\IEEEraisesectionheading{\section{Introduction}\label{sec:introduction}}

\IEEEPARstart{A}s end users are increasingly demanding higher performance from computer systems, processor vendors have been looking for every possible source of optimization and improvement in microarchitecture design.
Modern processors heavily rely on speculation to offer high instruction level parallelism. 
Under speculation, the processor executes instructions based on certain predicted paths, which may potentially resolve to be the wrong executions (i.e., transient execution). 
As transient execution can defy program software semantics, the underlying speculation engine is carefully designed to ensure that executions of undesired instructions are squashed, and no \emph{architectural state} changes are in effect for mis-speculation.

Recent advances in transient execution attacks have demonstrated the possibility of exploiting speculative execution to construct dangerous information leakage attacks. As the branch prediction unit (BPU) plays a key role in determining instructions to be fetched in the speculative path, it has been heavily exploited in these attacks to \emph{trigger mis-speculation}. Particularly, in Spectre V1, the attacker can induce speculative access of unintended data through branch direction mistraining, which enables the later leakage of secrets through a microarchitecture side channel~\cite{kocher2019spectre}. Although hardware-based side channels in the non-speculative domain are widely studied~\cite{wang2006covert,flush_reload,yao2018coherence,evtyushkin2018branchscope,yao2017covert}, transient execution attacks considerably empower these classical information leakage threats by expanding the attack surface to the speculative domain.
While several system-level mitigation techniques are proposed~\cite{retpoline_intel,stibp,ibpb}, recent studies show that these techniques either are ineffective towards certain attack variants or rarely employed in userspace due to performance concerns~\cite{SMoTherSpectre,evtyushkin2018branchscope}.

In this work, we demonstrate a new class of hardware-based information leakage that exploits BPU state updates within the speculative domain. Our key observation is that resolutions of conditional branch instruction in the \emph{speculative path} (e.g., nested speculation) alter the states of branch pattern history---the pattern history table (PHT) in particular. More critically, these \emph{speculative updates of PHT states} are not restored even after the squash of speculatively executed branches in modern processors. As branch instruction outcomes in the speculation domain can depend on data accessed in a domain beyond the programmer’s original intention, speculative branch execution can be potentially exploited to perform transient execution attacks in which the BPU is utilized as the \emph{secret transmitting hardware}. 
We systematically explore the aforementioned security vulnerability and implement a new form of BPU side/covert channel in the speculative domain, which we term \textbf{\mytitle}. Similar to how Spectre fuels traditional cache timing channels, \mytitle reveals a more severe security concern for branch predictors with the attack manifestation in the speculative world as compared to prior BPU side channels~\cite{evtyushkin2016understanding,huo2020bluethunder,evtyushkin2018branchscope}.
Furthermore, \mytitle exhibits two unique characteristics distinctive from existing speculation-based exploits: (i) \mytitle completely relies on the BPU for transient execution triggering and speculation-domain secret transmission, minimizing the hardware footprint for attackers. It can bypass the bulk of existing defense techniques mostly targeting protection in the cache hierarchy~\cite{invispec,saileshwar2019cleanupspec}; (ii) Different from Spectre attacks that depend on the relatively rare code gadget (e.g., memory access indirection~\cite{mambretti2019speculator}), \mytitle can utilize much simpler code patterns that are more commonly existing (e.g., branch whose conditional is based on a speculatively-loaded value). These enable even higher exploitability for \mytitle as the transient execution attack in real systems. 

This article considerably extends our prior work~\cite{branchspec} in the following aspects: i) We provide new insights about mode transitions and the PHT collision mechanism for hybrid branch predictor in commercial-off-the-shelf processors, which enable substantially higher efficiency in speculative secrets transmission. ii) We introduce a new variant of \mytitle side channel exploitation (i.e., \mytitle-v2) that chains arbitrary \mytitle gadgets through branch target poisoning, which further enhances the attack flexibility and capability over \mytitle-v1.  iii) We conduct a comprehensive investigation of code patterns in commonly-used application binaries to quantify the existence of \mytitle gadgets and demonstrate a real-world \mytitle attack against OpenSSL on Intel processors.
Our new findings further broaden the scope of the previous work and highlight the need for rethinking branch predictor designs that are secure in speculative executions. In summary, the key contributions are\footnote{Our PoC source is released at \url{http://tiny.cc/hfgutz}.}:
\textcolor{black}{
\begin{itemize}
\setlength\itemsep{0.2em}
    \item We find that speculative update of PHT states in modern processors creates a new information leakage threat that can leverage branch predictors as the transmitting medium in the speculative domain.
    \item We systematically explore the prediction mode transition in three recent generations of Intel processors and reverse-engineer the PHT collision mechanism under the history-based predictor. The discovery enables probing of the PHT state perturbations by an attacker with extremely high efficiency. 
    \item We present a novel transient execution attack framework--\emph{\mytitle}--that enables information leakage through BPU in various forms: \emph{\mytitle-cc}, an ultra-fast covert channel that achieves up to 1.3Mbps transmission bit rate; \emph{\mytitle-v1} and \emph{\mytitle-v2} side channels that leverage conditional branch mistraining and branch target poisoning respectively to induce speculative PHT update in nested speculation. 
    \item We perform extensive analysis on code bases of 10 popular open-source applications/libraries, which shows wider existence and stronger leakage capability of \mytitle gadgets in real systems. Further, our case study demonstrates a real-world \mytitle side channel on OpenSSL that achieves {97.3\%} bit accuracy. 
    \item We discuss potential speculation-secure branch predictor designs that can mitigate transient execution attacks exploiting modern branch predictors. 
\end{itemize}
}

\section{Background}
\label{sec:backgroud}

\subsection{Branch Prediction Structure}
\label{ss:branch_predictor}
Branch predictor is a critical per-core structure that directs the control flow of speculative execution. At a high level, the BPU is involved with two major tasks: \textit{direction prediction} that speculatively decides whether a conditional branch is taken or not, and \textit{destination prediction} that predicts branch target address. The BPU enables the processor to continue execution on the predicted path before the branch's outcome resolution to minimize pipeline stalls. The underlying speculation engine ensures that instructions will eventually retire in order using the re-order buffer and only committed instructions can change the architecturally visible states (e.g., architectural registers and memory).

\begin{figure}[t]
    \centering
    \includegraphics[width=0.37\textwidth]{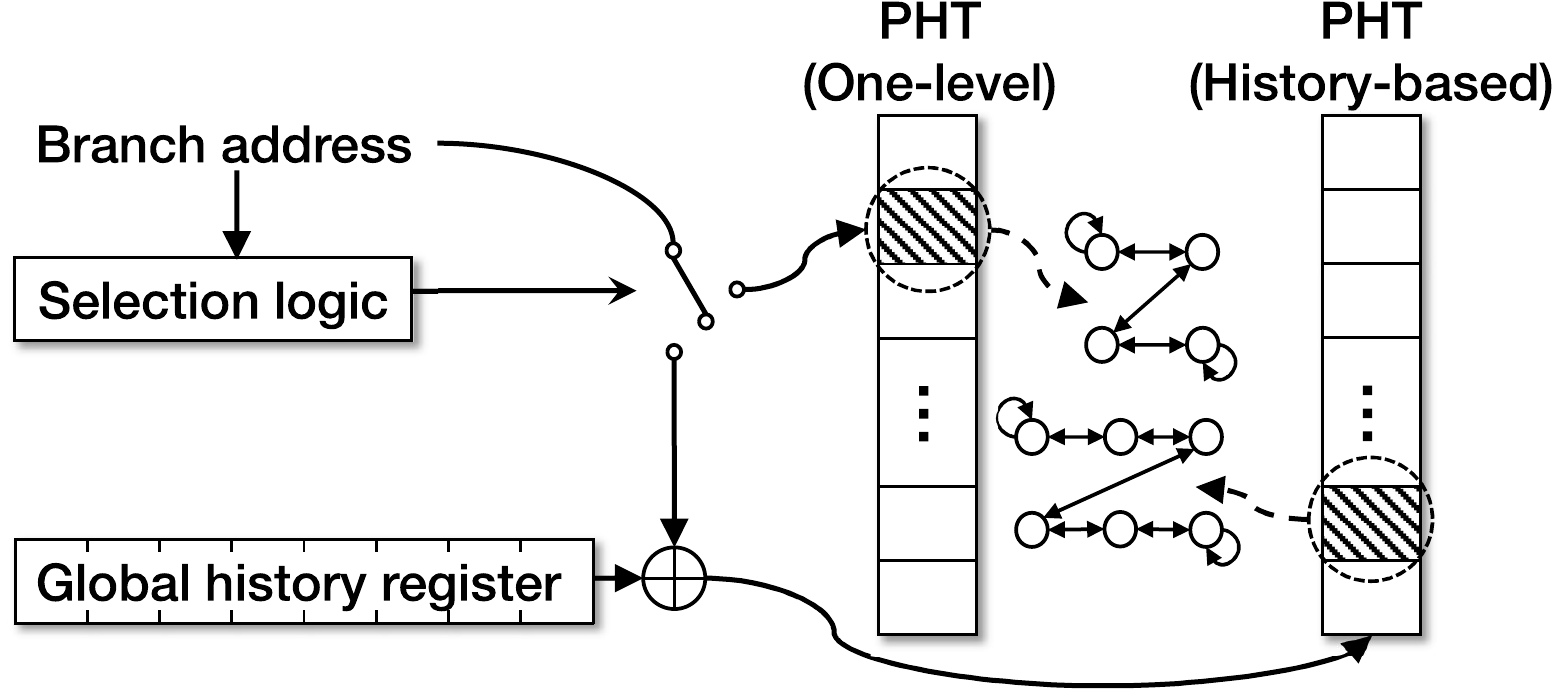}
    \caption{A high-level illustration of modern branch predictor.}
    \label{fig:bpu}
\end{figure}

Typically, the BPU takes advantage of the Pattern History Table (PHT) for direction prediction.
Each entry of the PHT incorporates a state machine using a saturating counter. For instance, in a 2-bit saturating counter, four possible states are associated with each PHT entry: \emph{Strongly Taken} (\textbf{ST}), \emph{Weakly Taken} (\textbf{WT}), \emph{Weakly Not Taken} (\textbf{WN}) and \emph{Strongly Not Taken} (\textbf{SN}). To predict branch directions, the BPU can operate in different modes~\cite{evtyushkin2018branchscope}. 
Specifically, in the one-level prediction mode, the branch address is the only information source to index the PHT entry. As a result, each branch only maps to a single PHT entry. One-level prediction excels in fast training, but it performs poorly for branches whose outcomes depend on the program execution context (e.g., the execution history of recently executed branches). Differently, history-based prediction (i.e., two-level prediction) maintains the history of prior branch executions in a branch history buffer. It leverages both the branch history and branch address to access the PHT and can train multiple PHT entries, each of which corresponds to a unique branching context~\cite{branchspec}. The history-based prediction mechanism can predict branches with complex patterns with substantially higher degree of accuracy at the expense of longer training time.

Modern processors (e.g., from Intel) generally use a hybrid design that leverages both one-level and history-based prediction in tournament mode as illustrated using \textit{Selection logic} in Figure~\ref{fig:bpu}. 
\textcolor{black}{In particular, the one-level prediction uses a predetermined number of bits from the branch address to select PHT entry. The history-based prediction instead combines the global history register (which represents the branch history state) with the branch address to index into PHT.} The global history register (GHR) is a shift register that keeps track of the most recent history among all branches executed on the core. Note that since the number of entries in PHT is limited, some branches will unavoidably map to the same PHT entry, leading to PHT collision that will create interference in predictions for those branches.

\vspace{-2mm}
\subsection{Microarchitectural Timing Channel Attacks}
Microarchitectural attacks are a class of information leakage threats where a malicious process manages to receive or infer secrets via a stealthy communication channel \emph{using microarchitectural components as the transmitting medium}. Among various attack variants, timing channels that modulate access latency to hardware resources are most widely exploited. These attacks can either manifest as covert channels that allow two isolated domains to willingly transmit data illegitimately or as side channels in which a spy process illicitly steals secrets from an unknowing victim process.  
Prior works have demonstrated timing channels on various hardware components in modern processors such as function units~\cite{wang2006covert}, caches~\cite{liu2016catalyst,kocher2019spectre}, and memory bus~\cite{venkataramani2016detecting}. Recent works~\cite{evtyushkin2018branchscope,huo2020bluethunder} show that attackers can infer \emph{program-defined secrets} that are used as branch conditionals by observing the perturbation in BPU microarchitectural states. To mitigate these classical timing channels, hardware-based techniques such as partitioning that avoid resource sharing~\cite{liu2016catalyst,yao2019cotsknight,yao2019leveraging}, obfuscating timing observations~\cite{fang2021defeating,yao2019covert} and randomizing hardware access patterns~\cite{qureshi2018ceaser} are proposed. 

\begin{figure}[t]
    \centering
    \includegraphics[width=0.455\textwidth]{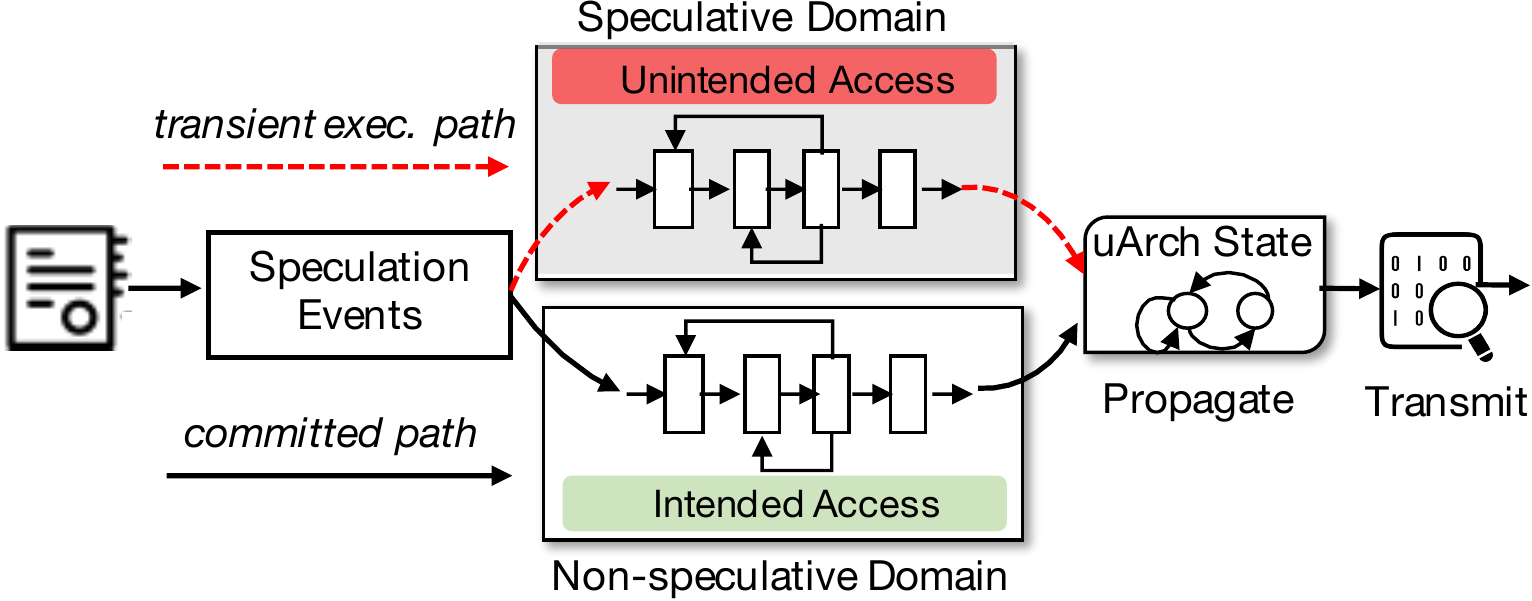}
    \caption{Side channels in the speculation and non-speculation domain.}
    \label{fig:speculative_execution} 
    \vspace{-0.1in}
\end{figure}

\subsection{Transient Execution Attacks}
\label{sec:bg-transient}
Transient execution attacks augment classical side channels by exploiting the effect of speculation in modern processors~\cite{kocher2019spectre}. They leverage the fact that the speculative execution path driven by the speculation engine can defy the program semantic in case of a branch misprediction. The tentative erroneous execution flow could lead to \emph{unintended memory accesses} that cross security boundary.  %
A successful transient execution attack depends on two factors: 1) the unintended accessed memory is propagated to taint certain microarchitectural states and 2) the microarchitectural states remain unpurged after mis-speculation is detected. \textcolor{black}{Figure~\ref{fig:speculative_execution} illustrates the high-level comparison between the transient execution attack and a non-speculative side channel.} While non-speculative side channels typically rely on the victim \emph{directly} using secret-dependent control flow or data flow, {transient execution attacks are even more dangerous as they substantially broaden the data outreach for an attacker}.

Spectre attacks abuse branch predictors to trigger transient execution of instructions that access restricted data. These attacks widely harness caches as the target hardware component for emitting secrets as memory blocks accessed by speculative loads/stores remain in cache even after speculation is rolled back. Particularly, the V1 variant mistrains the BPU to predict the wrong branch direction, while in V2, the attacker performs branch target injecting to hijack the speculative control flow. To mitigate spectre attacks, system-level defenses are employed that aim to either limit speculation through software patches (e.g., adding fences or using retpoline~\cite{retpoline_intel}) or restrain branch target poisoning (e.g., IBRS and STIBP~\cite{ibrs,stibp}). While these techniques can mitigate security breaches due to speculation, recent studies reveal that they either do not defeat all attack vectors or may introduce non-trivial performance overhead, which hinders its adoption in userspace applications~\cite{SMoTherSpectre,evtyushkin2018branchscope}.

\section{Threat Model}

Similar to previously demonstrated transient execution attacks~\cite{SMoTherSpectre,kocher2019spectre}, we assume the attacker can run a process on the same core with the victim. These two processes are running either on the same hardware context in a round-robin fashion or on individual virtual cores under simultaneous multi-threading. Since the PHT is a per-core structure, the victim's perturbations on the PHT can potentially be observed by the attacker. 
The attacker process only has userspace privileges, and the pre-requisite of a malicious OS is not required. Further, we assume that the attacker has knowledge about the code/binary of the victim application and can trigger victim's execution.

\section{PHT Update in Speculative Path}
\label{sec:speculative_pht}

Speculative execution can be triggered by a multitude of on- and off-chip events with varying resolution window sizes. For instance, speculation induced by contention in functional units may be resolved within only a few cycles, 
while it can take thousands of cycles if triggered by a last-level cache (LLC) miss followed by a row-buffer conflict in memory~\cite{mambretti2019speculator}. 
As branches are one of the most common instructions in programs, multiple branch instructions may be encountered in the speculation path of a program. To avoid pipeline stall in the front-end, modern processors perform \emph{nested speculation} where branch predictions continue to be made for branch instructions executed in the \emph{speculative path} within a certain speculation window~\cite{subramanian2017fractal}. 
In case the earlier branch that triggers the speculation resolves and a misprediction is detected, the processor will squash all dependent instructions - including the subsequent branches that are fetched along the speculative path. Note that if a speculatively executed branch is resolved before it is squashed, the processor can potentially update the branch predictor's states (e.g., PHT) based on its branch outcome.

\begin{lstlisting}[float,language=C, caption={Sample code with potential nested speculation.}, label={lst:pht_for_loop},belowskip=-15pt]
if (x < bound) // Outer branch
    // Inner branch in loop structure
    for (int i = 0; |\textbf{i < iterator}|; ++i)
        <some_operations>;
\end{lstlisting}

The code example from Listing~\ref{lst:pht_for_loop} shows why it may be beneficial to allow speculative PHT state updates. In this example, depending on the availability of \textit{bound} and \textit{iterator}, the inner branch (line 3) can be resolved very quickly. In contrast, the outer branch  (line 1) remains unresolved for several iterations of the inner branch. 
In case the outer branch is predicted as \emph{not taken} in the direction breaking out the loop, the inner branch may be executed for multiple iterations. Once the outer branch is resolved and the actual direction of the branch is known, the processor's states are rolled back. If the PHT entry for the inner branch is not updated in the speculative path, the BPU would not be trained properly for predicting the loop behavior. For instance, the inner branch could be mostly $taken$ while the initial PHT state for the inner branch is in the $not\ taken$ state. In this case, the inner branch will continue to be mispredicted, leading to degraded performance if the entire speculation path turns to be useful. 
In contrast, if the PHT is allowed to update in the speculative path, the PHT entry will converge to $taken$ after a few iterations of the inner branch. 
\begin{lstlisting}[float,language=C, caption={Testing PHT updates in speculative path.}, label={lst:sepculative_update},belowskip=-15pt]
bool control;
if (i < bound) {    // Parent branch (|\textcolor{darkgreen}{$b_p$}|)
    start = rdtsc();
    if (control)    // Child branch (|\textcolor{darkgreen}{$b_c$}|)
        <some_operations>;
    end = rdtsc();
}
\end{lstlisting}

Based on the above discussion, we can see that updating the PHT based on nested speculation can bring potential performance benefits. However, once the dependent branch is resolved and the validity of the entire speculation is refuted, the processor needs to decide how to deal with those speculative PHT updates. In particular, the processor may choose to restore the PHT to the states before speculation started. This can annul the impact of speculation with respect to the PHT perturbations by speculative branch executions in the wrong path. However, prior academic studies have shown that recovering speculative updates of the PHT when mis-speculation is detected brings negligible performance advantage~\cite{yale}. 
In order to figure out the PHT update mechanism in real processors, we design a microbenchmark that monitors the PHT perturbations for branch executions within mis-speculated paths. The core code snippet is shown in Listing~\ref{lst:sepculative_update} where a child branch $b_c$ can be speculatively executed when speculation is triggered by the parent branch $b_p$.
The experiment performs the following steps:

\vspace{1mm}
\noindent\textbf{\ding{202} Initialization:} In the first step, we execute a sequence of branch instructions with randomized outcomes~\cite{evtyushkin2018branchscope} that forces the BPU to use the one-level prediction (See Section~\ref{sec:reverse_engg} below for more discussions). 
The one-level prediction utilizes branch address exclusively to index PHT, thus making it easier to control collision in order to infer the state of a particular PHT entry later.

\vspace{1mm}
\noindent\textbf{\ding{203} Triggering $b_p$ misprediction:} In this step, we first train the $b_p$ branch with \emph{not taken} outcomes and subsequently trigger mis-speculation with an out-of-bound \texttt{i} value. 
Based on the value of $control$, $b_{c}$ will be \emph{taken} (or \emph{not taken}). The code segment in  Listing~\ref{lst:sepculative_update} is executed multiple times so that the PHT state of $b_{c}$ will converge to \emph{taken} (or \emph{not taken}) if BPU updates are not squashed for the transient executions of $b_{c}$. Note we load $control$ in cache and flush $bound$ out of cache to ensure $b_c$ is resolved before $b_p$.

\vspace{1mm}
\noindent\textbf{\ding{204} Infer the outcome of $b_c$:} In this step, we set an in-range value for $i$, and preload $i$ and $bound$ in cache. We then execute the code again with $control$ value set to 1 (i.e., $b_c$ should be \emph{taken}) and measure the latency for executing the code block in Line 4-5.

\begin{figure}[t]
	\centering
	\subfloat[b][Skylake]{
		\includegraphics[width=0.15\textwidth]{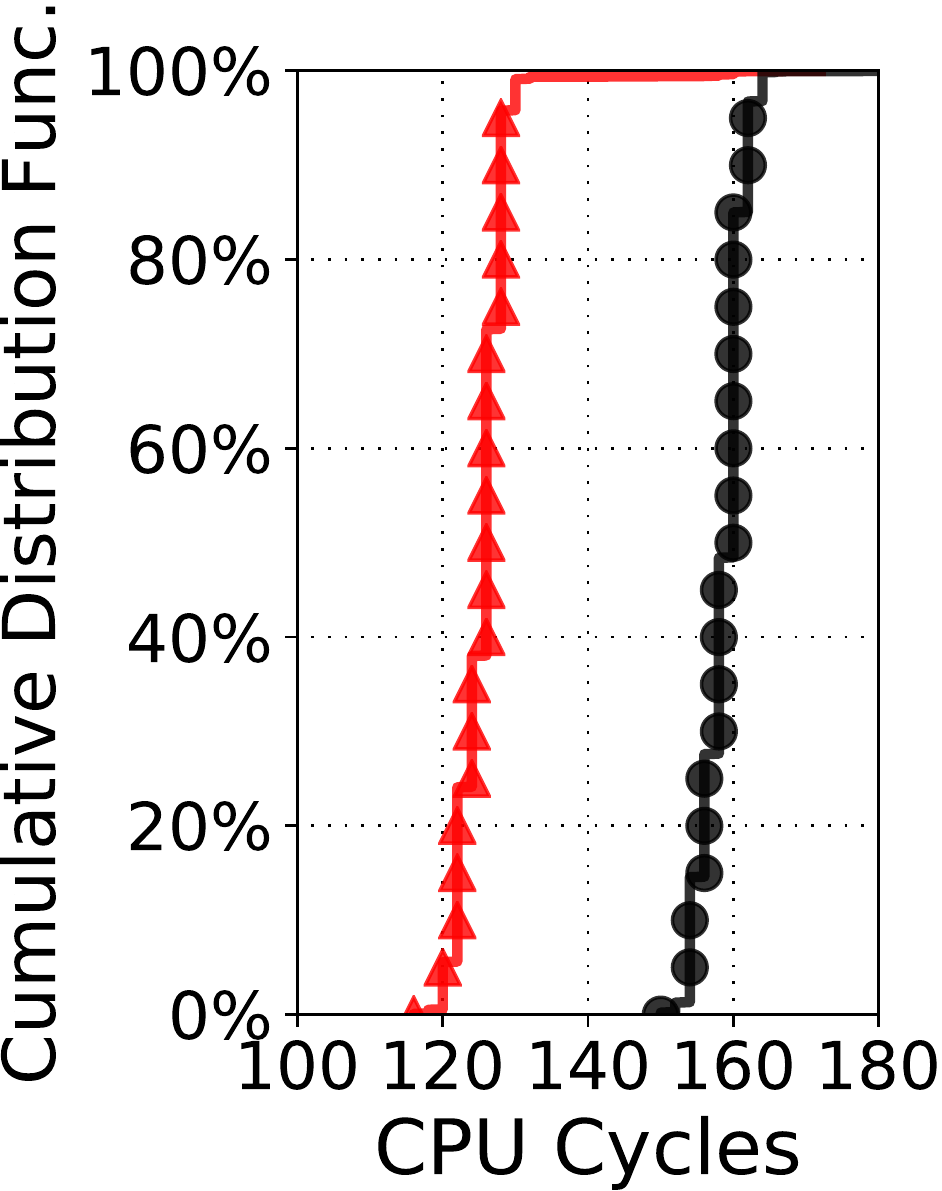}
		\label{fig:sky_latency}}
	\subfloat[b][Coffee Lake]{
		\includegraphics[width=0.15\textwidth]{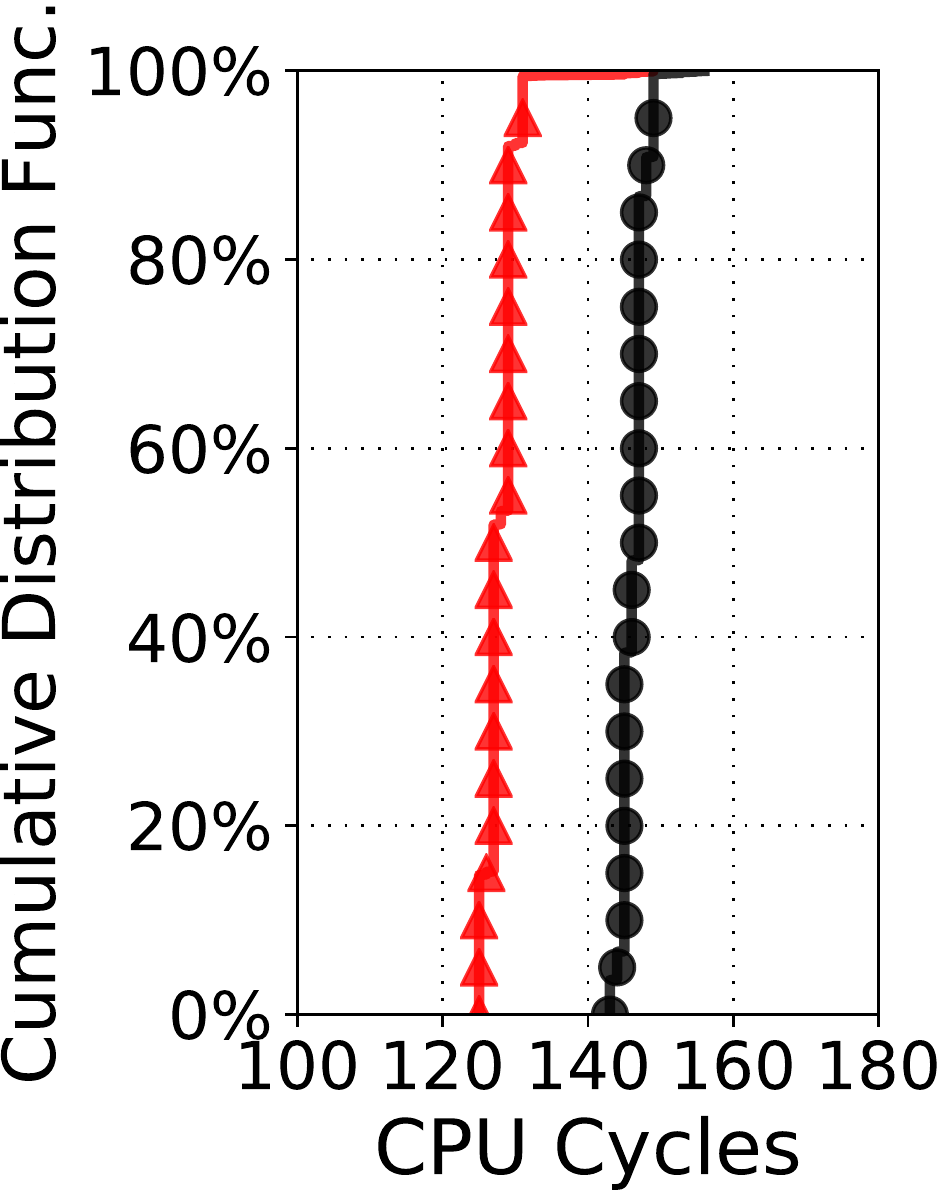}
		\label{fig:kaby_latency}}
	\subfloat[b][Cascade Lake]{
		\includegraphics[width=0.15\textwidth]{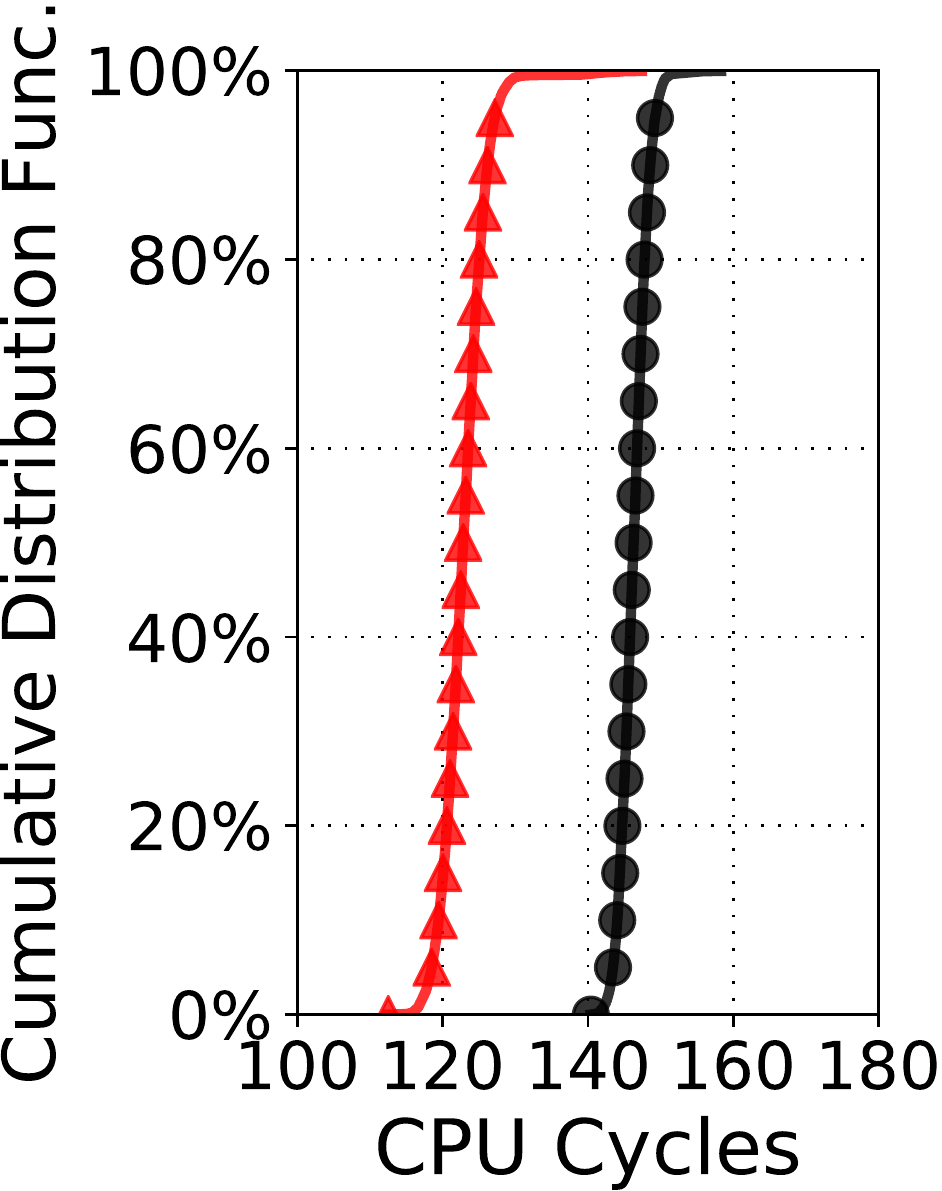}
		\label{fig:cascade_latency}}
	\caption{Execution latency distribution for $b_c$ branch in step \ding{204} corresponding to the \emph{Taken} and \emph{Not taken} outcome of $b_c$.}
	\label{fig:latency_test}
\end{figure}

We run this experiment on machines with three generations of Intel processors - \emph{Skylake}, \emph{Coffee Lake} and \emph{Cascade Lake}.
On each of the machines, we execute the aforementioned experiment 1000 times for each of the following configurations: 1) $b_c$ in step \ding{203} \emph{not taken}, and $b_c$ in step \ding{204} \emph{taken}; 2) $b_c$ in both step \ding{203} and \ding{204} are \emph{taken}. Note that this can be easily setup by controlling the value of $control$. In Figure~\ref{fig:latency_test}, we show the latency distributions for executing the code block line 4-5 in step \ding{204}. As we can clearly see, the execution time of the $b_c$ in step \ding{204} is consistently shorter (i.e., correct prediction) if the outcome of branch $b_c$ is the same as the \emph{speculatively resolved outcome} of that branch in \ding{203}. In contrast, when the outcome of $b_c$ in step \ding{204} is different from step \ding{203}, we observe longer execution time indicating a misprediction has occurred. These results evidently show that conditional branches resolved in transient execution do influence the branch prediction later on even after they are squashed. %
Based on the investigation results, we make the key observation that \textbf{conditional branches executed on speculative path changes the PHT state when the branch outcome is resolved, and these alterations are not restored regardless of whether the branch is eventually committed or not}. Moreover, such an observation is consistent across all processors we have evaluated. 
We note that the microarchitectural footprint in the BPU due to speculation can create a new avenue for transient execution attacks, essentially making it possible to exploit the BPU as the secret transmitting medium in the speculative domain. 
To the best of our knowledge, we are the first to investigate the exploitation of the PHT state changes in branch predictors for speculative branches in transient execution attacks. {Note that while we mainly explore the speculation behavior of BPUs in Intel processors, our observation could also be applicable to chips from other vendors. Particularly, any processor that allows speculative updates of the PHT can be vulnerable to this exploitation.}

\begin{figure}[t]
\centering
		\includegraphics[height=.4cm]{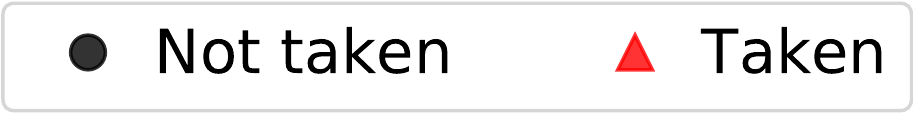}
		\vspace{-8mm}
\end{figure}

\section{Understanding Modern Branch Prediction Mechanisms}
\label{sec:reverse_engg}

\subsection{History-based Predictor Triggering Mechanism}
\label{ss:bpu_transition}
Understanding the branch prediction mode in operation and the conditions that trigger the BPU to transmit the prediction mode from one to another is critical for an attacker to create PHT collisions. Branchscope~\cite{evtyushkin2018branchscope} triggers the one-level predictor mode by running a large sequence of branches (more than 100K) with random outcomes {that cannot be well predicted by the history-based predictor.} 
While one-level prediction simplifies the procedure of PHT collision, it incurs substantial runtime overhead for the attacker as the randomization procedure is needed frequently. This is because one-level prediction is typically only active for a short period before the BPU resumes the history-based prediction mode that generally exhibits better prediction accuracy. 
A recent study~\cite{huo2020bluethunder} has proposed an extension of the BranchScope attack by exploiting the history-based predictor. It shows that history-based activation can be enforced by running an empirically found sequence of conditional branches with a certain length. However, the exact details of the triggering mechanism have not yet been fully explored. We systematically reverse-engineer the history-based prediction transitioning mechanism, enabling efficient controls of the prediction mode in the BPU.

\begin{figure}[t]
    \centering
    \includegraphics[width=0.48\textwidth]{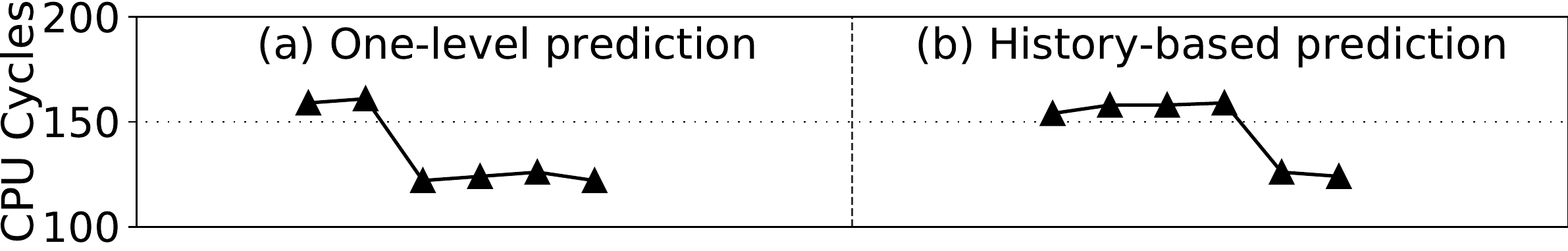}
    \caption{Pattern of prediction performance for a consecutive execution of target branch with \textit{Not Taken} outcome under different prediction mode. The PHT entry is initialized to \emph{strongly taken}. Longer latency indicates misprediction.}
    \label{fig:3-bit-pht}
\end{figure}

According to prior studies~\cite{huo2020bluethunder}, the saturating counters in the PHT have distinctive sizes in different prediction modes. Particularly, 2-bit counters (4 states) are used in the one-level prediction, and 3-bit ones (8 states) are utilized in the history-based prediction. For an $n$-bit saturating counter, values within [$0$, $2^{n-1}-1$] represent $taken$ states and [$2^{n-1}$, $2^{n}-1$] are the \emph{not taken} states (or vice versa).
If the branch predictor is first initialized with one-level prediction, it is possible to determine whether it has transitioned to the history-based prediction based on the misprediction behavior of a certain PHT entry. 
Specifically, when the target PHT entry for a branch is set to \emph{strongly taken}, executions of the same branch with \emph{not taken} outcomes (\texttt{NNN...N}) will result in 2 mispredictions in one-level prediction but 4 mispredictions in history-based prediction before they start to predict correctly. Based on this observation, we employ a two-step procedure to determine the current prediction mode: \emph{First}, a conditional branch is executed a sufficient number of times in one fixed direction (i.e., either \emph{taken} or \emph{not taken}), which trains the corresponding PHT entry to the \emph{strong} state (i.e., all '1's or '0's in the counter). \emph{Second}, the same branch is executed $K$ ($K>4$) times 
with the opposite branch outcome, and the execution latency of the basic block of the branch is measured. Note that while executing the same branch will guarantee the same PHT entry is accessed in one-level predictor, different PHT entries may be exercised for the same branch in history-based mode based on the GHR state. Therefore, to ensure only one PHT entry is used (in the case of history-based prediction), we run a sufficiently long sequence of predetermined branches before executing the target branch to preset the GHR. Figure~\ref{fig:3-bit-pht} shows the distinctive misprediction patterns that could be used to identify whether the current prediction mode is one-level and history-based. 

\begin{algorithm}[t]
\small
\DontPrintSemicolon

\SetKwInput{KwInput}{Input}                %
\SetKwInput{KwOutput}{Output} 

\KwInput{$t\_branch, seq\_length, mispred\_rates$}
\KwOutput{$pred\_mode$}

\For{$r$ $\in$ $\{mispred\_rates\}$}{
\tcp{Generate outcome sequence $S_r$ with expected misprediction rate $r$}
$S_r$ = gen\_seq (r, seq\_length)\\
\tcp{Execute t\_branch sequence with $S_r$ outcome}
exec (t\_branch, $S_r$)\\
\tcp{Check if history-based prediction is active}
pred\_mode = chk\_mode(t\_branch) 
}

\SetKwFunction{FMain}{chk\_mode}

\SetKwProg{Fn}{Function}{:}{}
\Fn{\FMain{t\_branch}}{
    \tcp{Set the test outcome sequence of the target branch}
    $S_t$ = \{\textbf{TTTTTTTT, NNNN}\}\\
    exec (t\_branch, $S_t$)\\
    \tcp{Check misprediction times (e.g., K set to 6)}
    {num\_mispred = mispredictions on last K executions}\\
    \If{num\_mispred == 4}{\KwRet \textit{History-based}}
    \Else{\KwRet \textit{One-level}}
}
\caption{Determining the triggering of history-based prediction}
\label{algo:two_level_transition}
\end{algorithm}

We hypothesize that modern processors employ a tournament-style design where the prediction accuracy of each prediction entity is dynamically monitored, and the best-performing prediction mode for each branch (or a set of branches) is selected.
To evaluate the transition criteria, we create a microbenchmark that executes a target branch instruction many times with \emph{a predetermined outcome}. This sequence can be configured such that the execution will result in a certain misprediction rate under the one-level prediction. We call such sequence the \emph{exercising sequence}. \textcolor{black}{
Note that the exercising sequence for a certain misprediction rate can be generated by first initializing the PHT entry state. For example, the branch sequence \texttt{TNTNTN} will lead to 50\% misprediction rate if the initial PHT state is set to \texttt{WN}}. 
After the exercising sequence is executed, the benchmark will test the current prediction mode by executing another instruction sequence with the same target branch - the \emph{testing sequence}. The branch outcomes during the testing sequence are set to distinguish the prediction patterns as shown in Figure~\ref{fig:3-bit-pht}.
The overall procedure is shown in Algorithm~\ref{algo:two_level_transition}.

\begin{figure}[t]
    \centering
    \includegraphics[width=0.38\textwidth]{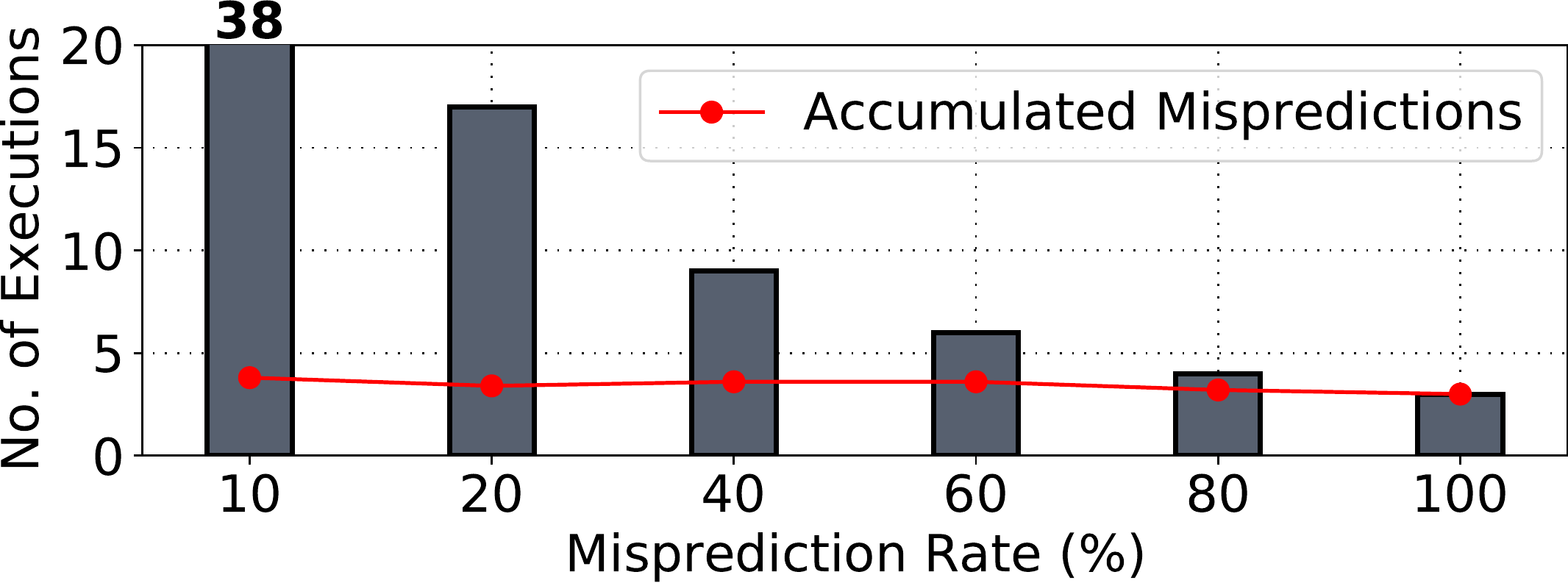}
    \caption{Minimum number of $target\_branch$ executions required to trigger history-based prediction under different misprediction rates.}
    \label{fig:mispredrate_min}
\end{figure}

For each exercising sequence at a specific misprediction rate (i.e., $S_r$), we execute the microbenchmark 100 times. The misprediction rate is confirmed through reading performance counters. 
At the end of each run, the program checks the current effective prediction mode. We then compute the success rate of enabling the history-based predictor. 
We perform the same experiment with varying lengths of the \emph{exercising sequence}. Figure~\ref{fig:mispredrate_min} illustrates the minimally required length under each aimed misprediction rate. The results reveal that the required number of executions for the target branch decreases as the misprediction rate increases accordingly. 
More clearly, we believe that the \textit{accumulated misprediction is used as the triggering criteria}. Particularly, when three mispredictions occur under the one-level prediction, transitioning to the history-based prediction happens. We note that such phenomenon can be potentially attributed to the internal selection logic that uses a confidence counter to choose a winning prediction mode~\cite{kessler1999alpha}. Using this knowledge, we can swiftly trigger the history-based predictor by only executing the target branch six times with the corresponding outcomes: \texttt{TNTNTN}. {Note that this sequence could either induce three or six mispredictions for one-level prediction based on the initial state of the PHT entry.}

\begin{figure}[t]
    \centering
    \includegraphics[width=0.42\textwidth]{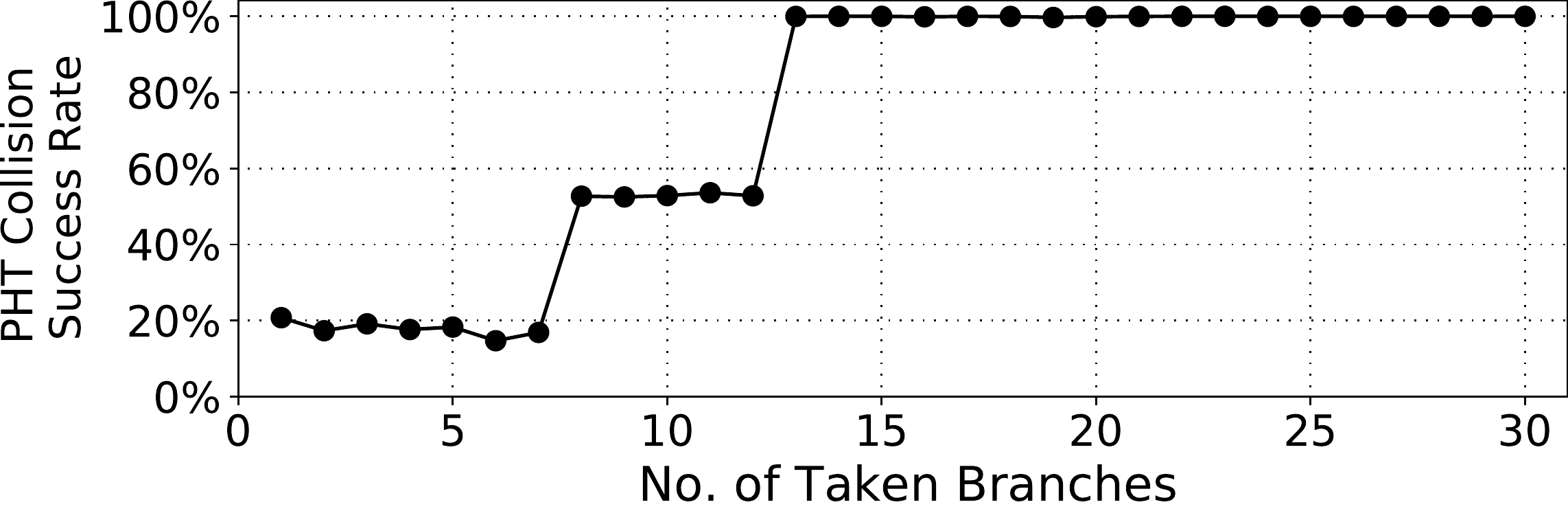}
    \caption{Number of taken branches executed to create PHT collision under history-based predictor.}
    \label{fig:taken_branch}
\end{figure}

\subsection{Creating PHT Collisions in History-based Predictor}
\label{ss:ghr_preset}
With the history-based prediction, the PHT entry of a conditional branch depends on the state of branch pattern history stored in the GHR. As a result, multiple PHT entries may be trained for predicting one branch in this mode. To create a PHT collision in the history-based prediction, using a congruent branch (as in one-level prediction) is no longer sufficient. Particularly, the GHR has to be properly set for both the observee and observer branches (i.e., the branch of the victim and the attacker in side channel).  
One possible way to do this is to execute an excessive number of conditional branches to ensure the GHR is flushed. However, such a mechanism undermines the efficiency of PHT state inference. A more optimized approach is to precisely configure the GHR, which requires knowledge about the size of the GHR and how it is populated. Classical BPU design implements the GHR as a shift register where '1' or '0' are inserted when a conditional branch is resolved as \emph{taken} and \emph{not taken} respectively~\cite{gshare}. However, prior studies have shown that the GHR in Intel processors is populated with \emph{partial bits} from the target addresses of \emph{taken branches}~\cite{google-zero}.  
\emph{To determine the exact size of the GHR}, the following experiment is performed: 1) activate the history-based prediction for the target branch (as discussed in Section~\ref{ss:bpu_transition}), 2) preset certain PHT entry to \emph{strongly taken} by executing \texttt{$N$} distinctive $taken$ branches followed by a target branch with $taken$ outcome, 3) detect PHT collision by executing the same \texttt{$N$} number of $taken$ branches followed by the execution of target branch with $not\ taken$ outcome. %

We vary \texttt{$N$} and detect PHT collision accuracy under each setting.
This experiment is run 1000 times for each \texttt{$N$} value and the results are shown in Figure~\ref{fig:taken_branch}. We can see that \textbf{executing 12 taken branches before the target branch} is sufficient to preset the state of GHR for direction prediction and ensure PHT entry collision. Such observation is consistent among all processors we tested. 
We therefore conjecture that the size of GHR used for history-based prediction is $12\times B_t$ where $B_t$ is the number of bits from the targeted address of a taken branch populated to the GHR.

\section{Overview of Exploitation}
\label{sec:overview_branchspec}

In this section, we show the overview of the \mytitle attack design which performs information leakage by inferring secrets from BPU state updates in transient execution.

As discussed in Section~\ref{sec:reverse_engg}, for a PHT entry with an $n$-bit saturating counter with $2^n$ possible values, there are 1 \emph{Strongly Taken} state, $2^{n-1}-1$ \emph{Weakly Taken} states, $2^{n-1}-1$ \emph{Weakly Not-taken} state and 1 \emph{Strongly Not-taken} state.
Once a PHT collision is achieved, the attacker can infer secrets by observing the sequence of prediction outcomes made for a colliding branch. %
Specifically, to infer the outcome of a victim branch $b_v$ when \emph{executed speculatively}, we first use a colliding branch $b_a$ from the attacker's address space\footnote{\label{footnote_1}Note that from this point, executing a target branch in the history-based predictor will implicitly mean that GHR is properly preset to ensure collision.} to initialize the target PHT entry to a strong state. The attacker then triggers the execution of a victim's branch $b_v$ speculatively. Finally, we execute $b_a$ again to infer the state of the PHT that has already been perturbed by the victim. If the outcome of $b_v$ is dependent on a secretive value (i.e., unintended secret), the attacker can reveal that value after the mis-speculation is corrected. 
Figure~\ref{fig:general_attack} shows a high-level implementation of the attack. Note that while we use side channel terminologies in this description, the same techniques can also be applied to covert channels. 
We now discuss how the attackers achieve each of these steps:

\begin{figure}[t]
    \centering
    \includegraphics[width=0.485\textwidth]{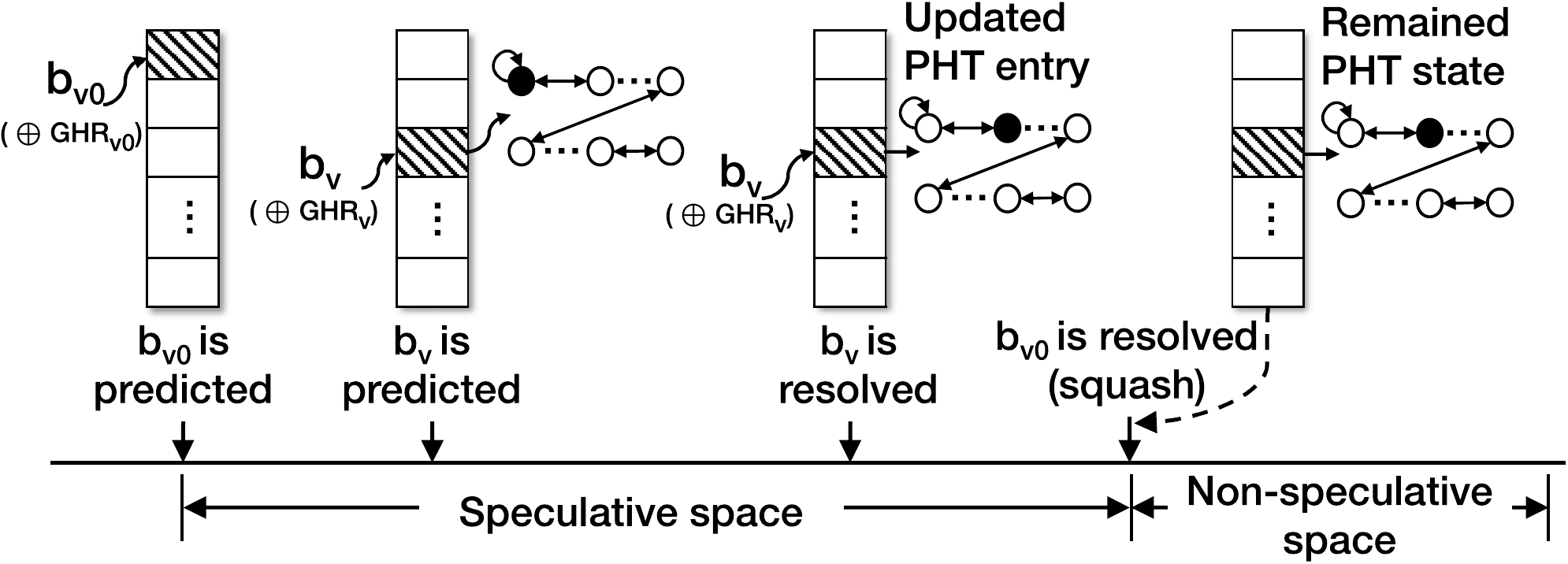}
    \caption{Illustration of information leakage through speculative branch executions. 
    }
    \label{fig:general_attack}
\end{figure}

\begin{enumerate}[label=\textbf{Step \arabic{enumi}: }, leftmargin=0cm,itemindent=.5cm,labelwidth=\itemindent,labelsep=0cm,align=left,itemsep=2pt]

\item \textbf{PHT initialization for victim's branch.} 
The attacker has two goals in this stage. First, the attacker trains the PHT entry of the victim's branch ($PHT_t$) so that it is pushed to a deterministic state (e.g., either \emph{ST} or \emph{SN}). 
For $n$-bit counters, this can be achieved by executing a branch $b_a$ in the attacker’s address space that is congruent to $b_v$ for $2^{n}-1$ times with the \emph{taken} (or \emph{not taken}) outcome. This means executions of $b_a$ $3$ and $7$ times for one-level and history-based predictor respectively. 
Second, the attack either mistrains the branch direction prediction (in case of a conditional branch) or poisons the target address (in case of an indirect jump) of the parent instruction $b_{v0}$ in the victim's process.
This will trigger transient execution of $b_{v0}$ in the path with the $b_{v}$ instruction. Finally, the attacker triggers the victim process execution and waits until $b_v$ is executed speculatively.

\item \textbf{Victim execution in speculative path.} 
When the victim runs, branch $b_v$ will be first speculatively executed and later squashed.
The attacker can control the speculation window for $b_{v0}$ to make the speculation sufficiently long so that $b_v$ is resolved first in the speculative path. $b_v$'s speculative resolution will alter the state of $PHT_t$ depending on certain conditionals (likely unintended data). 
Essentially, after victim’s execution, $PHT_t$ is tainted with the out-of-bound value. 
    
\item \textbf{Infer secret by probing $\mathbf{PHT_t}$ state.} 
In this step, the attacker aims to infer the secret value in the victim’s address space by probing the state of $PHT_t$. To do so, the attacker executes the branch $b_a$ that is congruent to $b_v$ with outcome in the \textit{opposite direction} from \textit{Step 1}. To observe the difference, the attacker executes $b_a$ for $2^{n-1}$ times and record the execution latency of $b_a$ execution.

\end{enumerate}

\begin{figure}[t]
	\centering
	\includegraphics[width=0.465\textwidth]{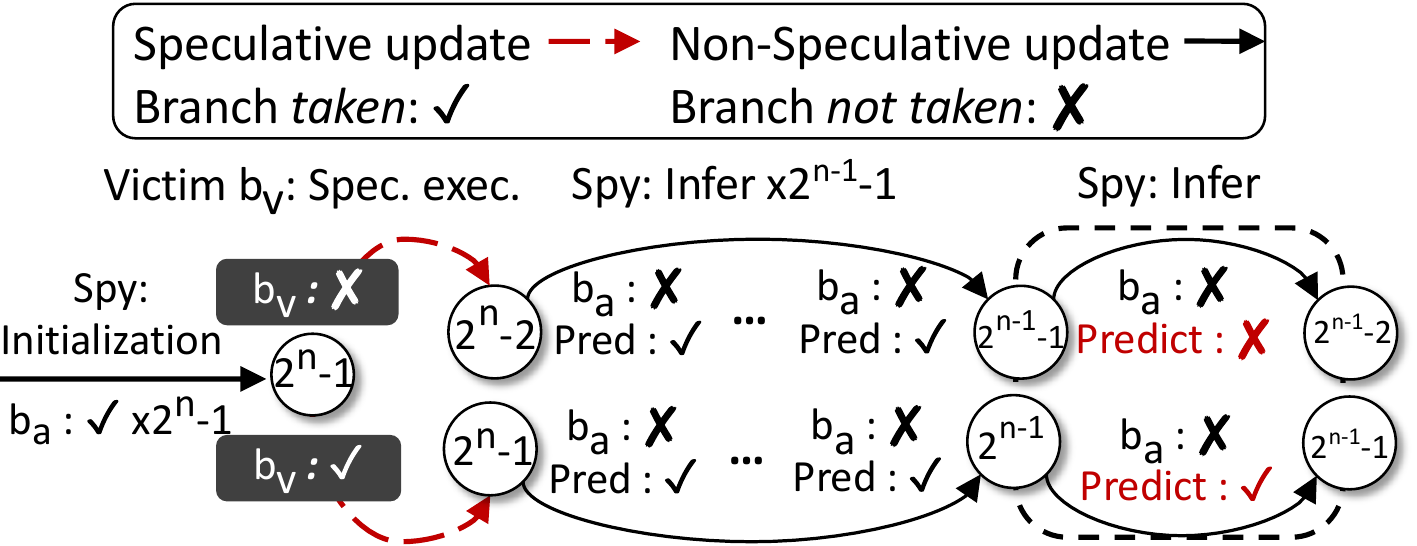}
	\label{fig:pht_1}
	\caption{Possible changes of $PHT_t$ states under exploitation (for $n$-bit saturating counters). The circled values denote states of the saturating counters at each stage.}
	\label{fig:pht_branchspec_overview}
\end{figure}

Figure~\ref{fig:pht_branchspec_overview} demonstrates the generalized state transition of $PHT_t$ for $n$-bit counters after the attacker presets $PHT_t$ state to \emph{taken} in Step 1. We can see that the value of $secret$ is directly correlated with the prediction of the $2^{n-1} th$ inference operation in Step 3. Particularly, if $b_v$ is resolved as \emph{not taken} speculatively, the $2^{n-1} th$ branch of the attacker will be correctly predicted, otherwise, a misprediction would occur. The attacker can then infer the secret used as $b_v$'s conditional based on timing as shown in Figure~\ref{fig:latency_test}.
Similar PHT state diagrams can be generated for branches with \emph{not taken} outcomes in Step 1 and \emph{taken} outcomes in Step 3.

\begin{figure}[t]
    \centering
    \includegraphics[width=0.37\textwidth]{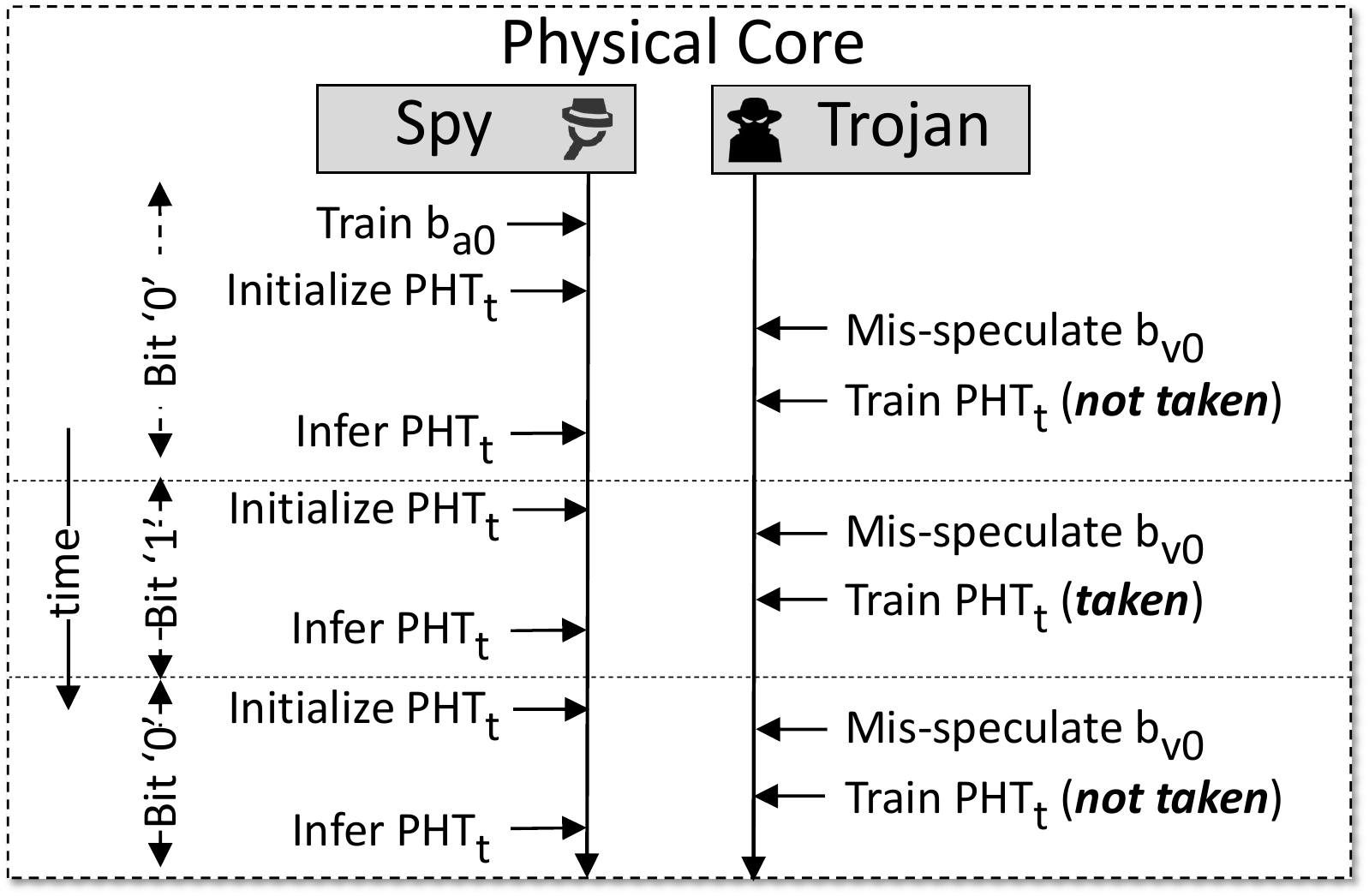}
    \caption{Illustration of \mytitle covert channel protocol.}
    \label{fig:protocol}

\end{figure}

\begin{figure}[t]
	\centering
	\includegraphics[width=0.46\textwidth]{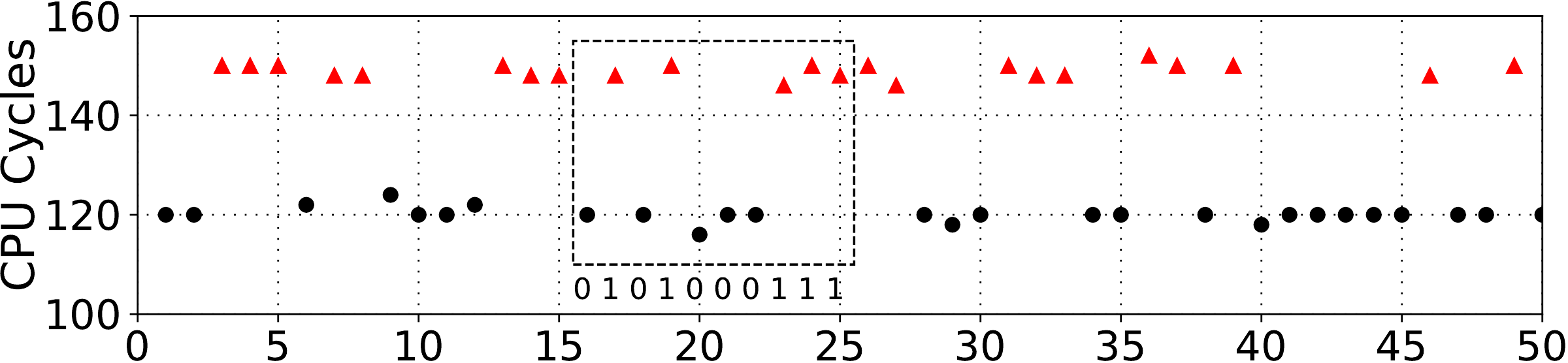}
	\caption{Latency traces for a 50-bit transmission by \textit{Spy} corresponding to the covert channel in Figure~\ref{fig:protocol}. Decoding for bit 16-25 is highlighted in the boxes.}
	\label{fig:covert_channel_demo}
\end{figure}

\section{\mytitle Covert Channel Attack}
To demonstrate the information leakage threat with speculative PHT update, we investigate a covert channel where a trojan and a spy exploit transient branch execution to build a covert communication. We call it \emph{\mytitle-cc}. Different from previous covert channel attacks in the non-speculative domain~\cite{huo2020bluethunder,evtyushkin2016understanding,evtyushkin2018branchscope}, covert channels using speculation can be more stealthy and remain undetected even with the presence of dynamic software analysis techniques~\cite{wang2017cached}.

To construct the \mytitle-cc attack, the \textit{spy} first executes Step 1 from Section~\ref{sec:overview_branchspec} and then waits for the \textit{trojan}'s execution (Step 2) before inferring the secret in Step 3. The code gadget for trojan's exploitation can be any value-dependent conditional branch executed in the speculative path. After activating certain branch prediction mode, the attacker needs to follow this sequence of actions: \textit{Spy initialization--}preset the $PHT_t$ to the deterministic state by executing $b_a$ $\rightarrow$ \textit{Trojan training--}execute the $b_v$ speculatively with conditional depending on sensitive data $\rightarrow$ \textit{Spy inference--}infer the state of $PHT_t$ after trojan's execution. \textcolor{black}{Since the trojan and spy are colluding, a PHT collision between them can be achieved by simply using branches (both $b_a$ and $b_v$) with the same address (for one-level prediction) or by executing the same set of 12 taken branches to preset the GHR state before the execution of $b_a$ and $b_v$ (for history-based prediction).} Figure~\ref{fig:protocol} shows the communication protocol of the covert channel and illustrates how the trojan transmits bits '010'. Figure~\ref{fig:covert_channel_demo} illustrates the latency traces observed by the \textit{spy} corresponding to the $2^{n-1}$th execution in the inference phase for a snippet of 50-bit transmission. The spy observes a clear pattern differentiating bit '0' and bit '1'. 

\begin{figure}[t]
	\centering
	\subfloat[b][One-level prediction]{
		\includegraphics[width=0.22\textwidth]{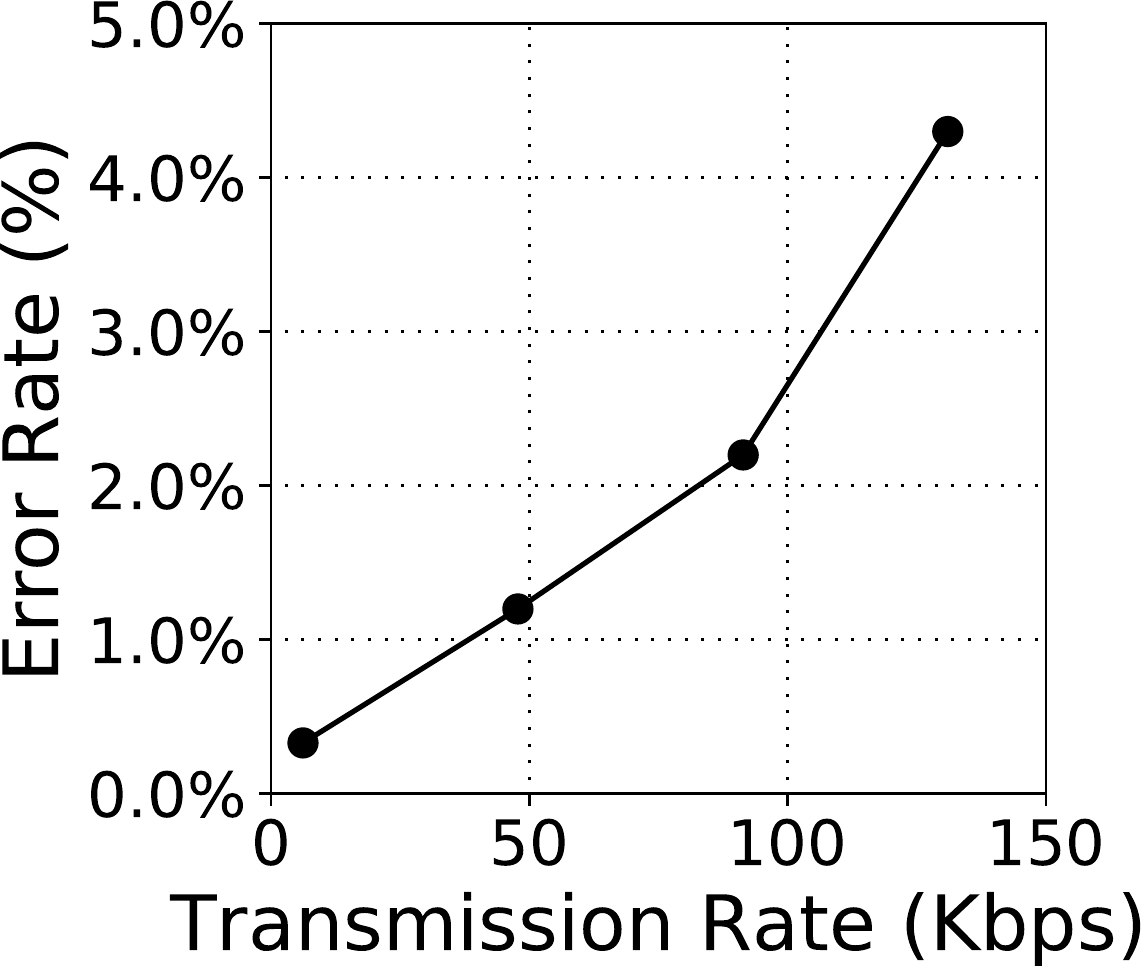}
		\label{fig:t_rate_1}}
	\subfloat[b][History-based prediction]{
		\includegraphics[width=0.22\textwidth]{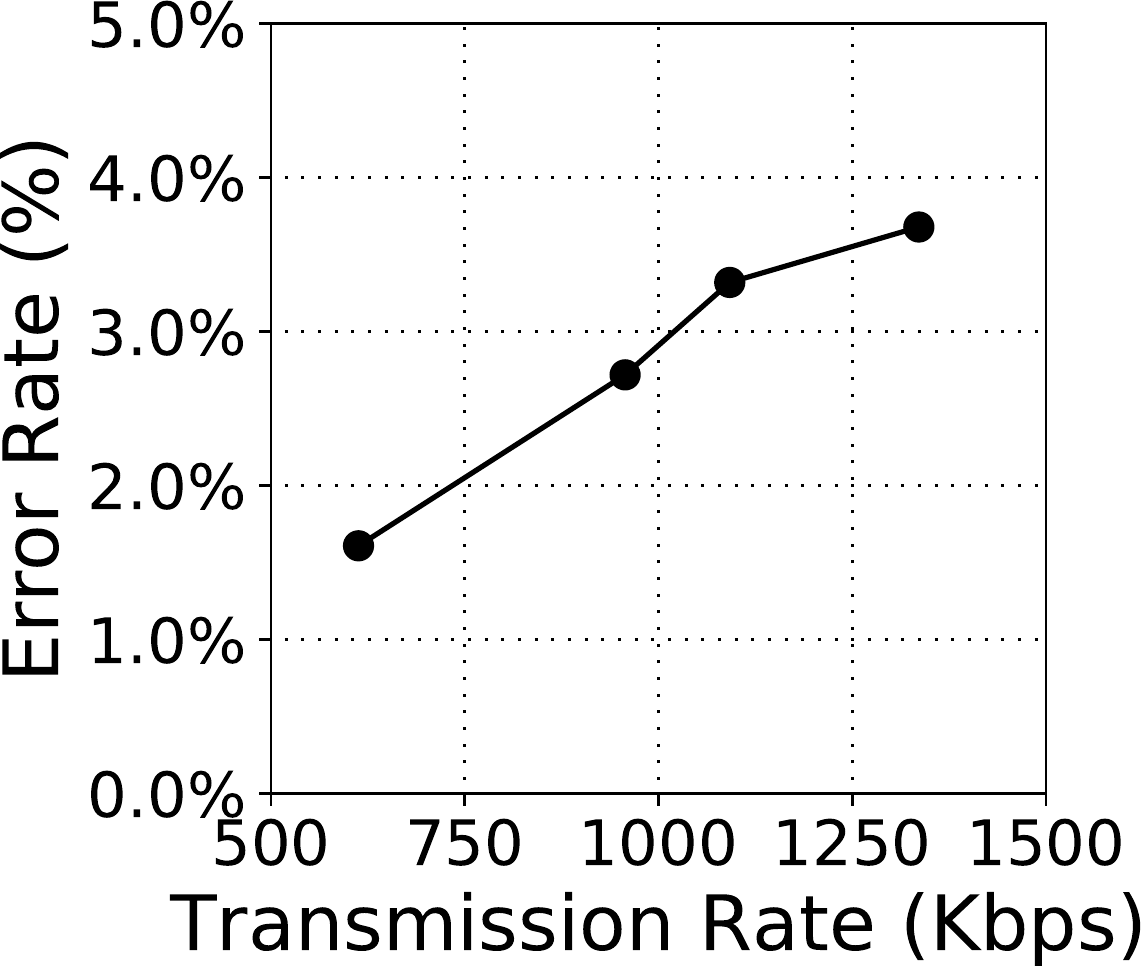}
		\label{fig:t_rate_2}}
	\caption{Accuracy of the covert communication channel as a function of transmission rates.}
	\label{fig:t_rate}
\end{figure}

\begin{figure*}[t]
	\centering
	\includegraphics[width=0.85\textwidth]{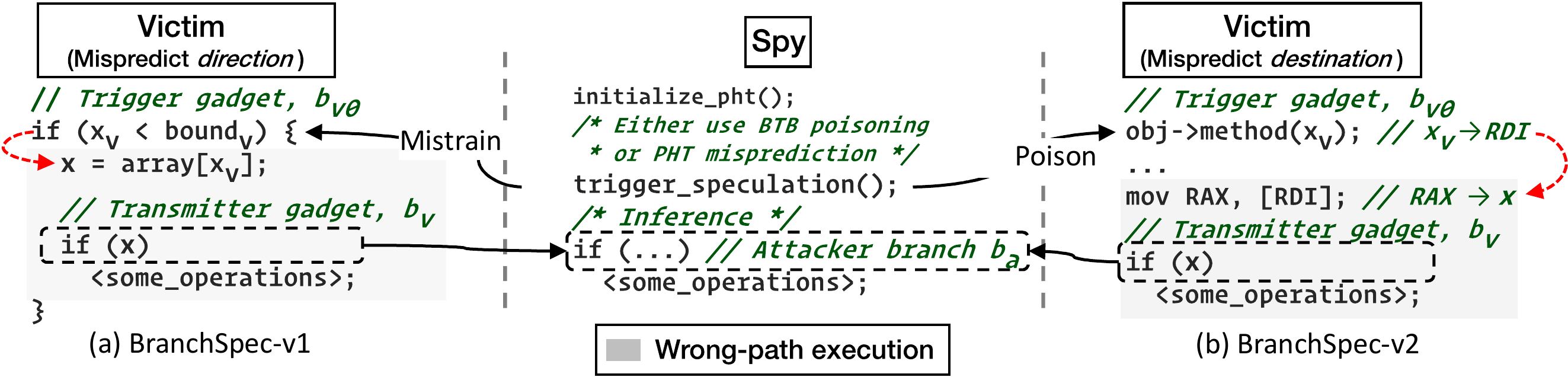}
	\vspace{-2mm}
	\caption{High-level implementation of \mytitle side channels to infer \textit{speculative changes} of PHT states. Here, (a) \texttt{x} can be unintended secret while \texttt{array} itself is not secretive in regular path of program execution or (b) \texttt{x} can be tainted by unintended memory access referenced by register \texttt{RDI}.}
	\vspace{-2mm}
	\label{fig:side_attack_overview}
\end{figure*}

There are several ways the transmission rate of the covert channel can be improved.
A common optimization technique is to reuse the operation in the inference phase for one-bit transmission as initialization operation for the next bit. Specifically, we increase the number of branch executions in the inference phase to $2^n-1$. This way, at the end of the inference, the target PHT entry is already set to a strong state (the purpose of the initialization step for the next bit reception). Note that the spy infers the secret by observing the prediction from $2^{n-1}$th inference branch. We can thus improve the bit rate by removing spy's \emph{initialization} through alternating the PHT entry of $b_a$ between $ST$ and $SN$.
We also perform speed-enhancing techniques particular to certain prediction mode.
For instance, under one-level prediction, we can increase the bit rate by tuning bits transmitted between re-enforcements of the one-level predictor.

Figure~\ref{fig:t_rate} illustrates the raw transmission rate of \mytitle-cc and the corresponding error rate under each prediction mode. Specifically, with one-level predictor, it is observed the dominant bit rate improvement is due to coalescing multiple bits transmission for one randomization operation. 
The attacker can achieve a peak transmission rate of 131Kbps within 5\% bit error rate (shown in Figure~\ref{fig:t_rate_1}). We find that further increasing bit rate will lead to considerable drop in bit accuracy due to transfer of the prediction mode. On the other hand, 
Figure~\ref{fig:t_rate_2} shows that under history-based prediction, the adversary can achieve up to 1.3Mbps with less than 4\% bit error ratio, which is an order of magnitude faster than the one in one-level prediction mode. We note that \mytitle-cc with history-based predictor is considerably more efficient since it eliminates expensive additional operations for keeping the one-level predictor active. 
\textcolor{black}{
Compared to the existing BPU-based covert channel leveraging coarse-grained pattern history manipulations~\cite{evtyushkin2016understanding}, \mytitle-cc exhibits much higher transmission rate due to precise control of collision on a single PHT entry in history-based prediction mode.
}

\section{\mytitle Side Channel Attack}
In this section, we demonstrate \mytitle side channels that enable inferring speculation-domain secrets from a victim process through branch predictor exploitation. To leverage this vulnerability, the attacker needs to find appropriate gadgets in the victim application. Particularly, \mytitle depends on the presence of two types of gadgets in the victim application: a \textit{trigger gadget} to start mis-speculation and a \textit{transmitter gadget} that perturbs a target PHT entry with the information of \emph{speculative secret}.

\vspace{1mm}
\noindent\textbf{Triggering gadget.} 
Generally, any speculation inducing instruction that deviates control flow can be a trigger gadget. However, to be \textit{exploitable}, the triggering gadget needs to fulfill two goals: (i) speculative execution in the wrong path driving towards the transmitter gadget and (ii) preparation of speculatively accessed secret (e.g., propagating the secrets to the instruction operands in the transmitter).

\vspace{1mm}
\noindent\textbf{Transmitter gadget.} The transmitter gadget can be as simple as a conditional branch that uses out-of-bound accessed data or more generally a register/memory (tainted by a secret in the speculative domain) as a part of conditional argument. Note that one such conditional jump using the tainted register is sufficient as this will alter the state of PHT for that branch according to the secret value. 

Our proposed attack can manifest in ways similar to either Spectre V1 or V2, and we call our attacks \emph{\mytitle-v1} and \emph{\mytitle-v2} respectively. The \mytitle-v1 attack leverages a conditional branch to induce mis-speculation. Its speculative execution path that traverses the trigger gadget and transmitter gadget follows through the static control flow graph of the victim program. Differently, \mytitle-v2 harnesses branch target positioning as the triggering mechanism. The exploited path is driven by chaining the attack gadgets at potentially arbitrary locations. As a result, the speculative path in \mytitle-v2 is not constrained by the victim's static control flow. Since the execution path in v2-type attack can manipulate instruction sequences throughout the entire address space, it provides the adversaries with higher attack flexibility as well as code gadget availability.

\subsection{Side Channel Implementation}
\label{ss:side_channl_impl}

Using the attack methodology discussed in Section~\ref{sec:overview_branchspec}, we can now build the two variants of \mytitle side channels. 
Specifically, the adversary locates a code sequence that corresponds to a transient execution path covering both a trigger gadget and a transmitter gadget. For \mytitle-v1, the code sequence contains two conditional branches where the first branch (e.g., \texttt{CMP <conditions>}$\rightarrow$\texttt{Condtional Jump <LABEL>}) 
induces mis-speculation that leads to the execution of the second one with \emph{speculative conditional values}. In \mytitle-v2, the trigger gadget ends with an indirect jump/call (e.g., a virtual function invocation), and its target address will be pointing to the transmitter gadget (i.e., speculative conditional branch) through branch target buffer (BTB) poisoning. Figure~\ref{fig:side_attack_overview} illustrates the primary steps of side channel exploiting for \mytitle-v1 and \mytitle-v2 variants. 
Once the victim branch is identified and its corresponding PHT entry is determined, the attacker locates a branch in its own address space that will collide with the same PHT entry ($PHT_t$). As shown in Figure~\ref{fig:side_attack_overview}, $PHT_t$ is first initialized by the attacker to a predetermined state. The attacker then triggers transient execution of the victim's target branch $b_v$ in the speculative path, which resolves before its earlier dependent branch is squashed. Lastly, the attacker infers the secrets by observing $PHT_t$ state changes made by the speculative victim branch using the technique shown in Section~\ref{sec:overview_branchspec}.

\vspace{1mm}
\noindent\textbf{Achieving PHT collision in side channels.}
\textcolor{black}{Under the \emph{one-level predictor}, the attacker can achieve PHT collision by using an attack branch ($b_a$) with the same or congruent address as the victim branch $b_v$.
For \emph{history-based prediction}, the attacker has to ensure that the GHR state (filled by the last 12 taken branches) before the execution of $b_a$ is exactly the same as the one before the execution of $b_v$ in the victim process. %
With a privileged attacker, this can be achieved by interrupting the victim before $b_v$'s execution and preparing a predetermined GHR value by the attacker. 
However, it is potentially more challenging for unprivileged attackers since they cannot control context switch of the victim arbitrarily. Fortunately, we observe that it is not uncommon that branches leading to $b_v$ are secret independent and they exhibit persistent prediction behaviors.    
As a result, it is possible for the attacker to perform off-line profiling of the victim binary with sample inputs to replicate the branch history in its inference stage. 
\textcolor{black}{
To ensure collision, gadgets exploitable for history-based prediction have the additional constraint that sufficient deterministic conditional branches preceding the transmitter gadget exist. Note that if most of the branches (not all) that impact the GHR state during the execution of the transmitter gadget can be determined, the attacker may still observe the perturbation in $PHT_t$ by $b_v$ by probing all possible PHT entries that would be touched.}
}

\begin{table}[t]
\centering
\begin{tabular}[t]{m{0.4\columnwidth}|ll}
\textbf{(a) Attacker} & &  \textbf{(b) Victim} \\
\midrule
\begin{lstlisting}[frame=none,xleftmargin=-3pt,framexleftmargin=0pt,aboveskip=0pt,belowskip=0pt,numbers=left,
    stepnumber=1]
array1[] = [];

main() {
 initialize_pht();
 /* Trigger victim */
 /* Infer */
 start = rdtsc();
 preset_ghr();
 if (|\emph{outer\_branch}|) {
  if (|\emph{inner\_branch}|) // |\textcolor{darkgreen}{\textbf{$b_{a}$}}|
   <some_operations>;
 }
 start = rdtsc();
}
\end{lstlisting} & &
\begin{lstlisting}[frame=none,numbers=left]
sec[] = {1,1,0,1,1,
        1,0,0,0,1};
array1[] = {0};

main(char *argv[]) {
 x = atoi(argv[1]);
 preset_ghr();
 // Trigger gadget: |\textcolor{darkgreen}{\textbf{$b_{v0}$}}|
  if (|$x$| < |$bound$|) {
    // Transmitter gadget: |\textcolor{darkgreen}{\textbf{$b_{v}$}}|
    if (array1[x])
      sum++;
    else sum|-\/-|;
  }
}
\end{lstlisting}
\\
\bottomrule
\end{tabular}
\vspace{1mm}
\begin{lstlisting}[frame=none,caption={Overview of of \mytitle side channel PoC. \texttt{Infer} step will be executed multiple times depending on which prediction mode is used.},label={lst:overview},aboveskip=0pt,belowskip=0pt]
\end{lstlisting}
\end{table}

\begin{figure}[t]
	\centering
    \subfloat[b][One-level Predictor]{
	\includegraphics[width=0.42\textwidth]{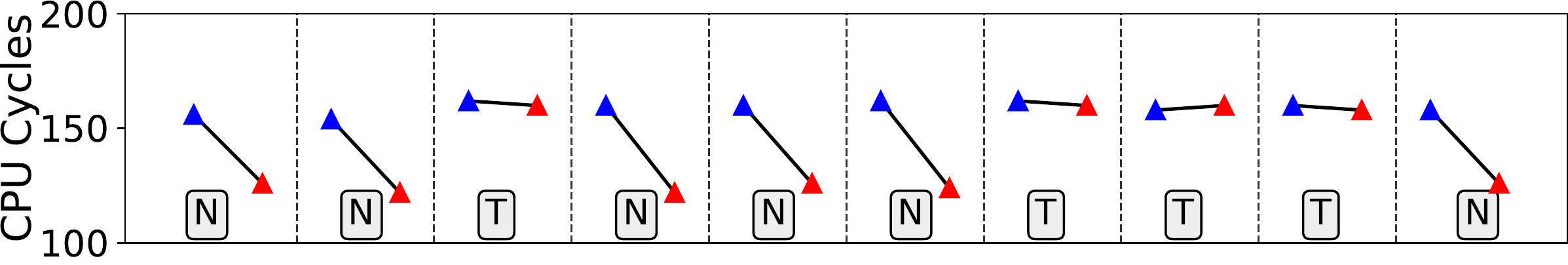}
	\label{fig:side_channel_one}}

	\subfloat[b][History-based Predictor]{
	\includegraphics[width=0.42\textwidth]{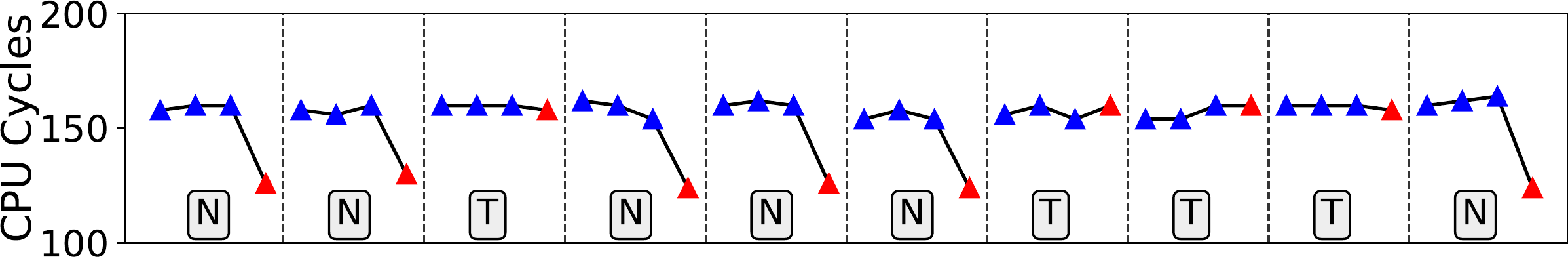}
	\label{fig:side_channel_two}}
	\caption{Side channel attack to recover the direction of victim branch execution (shown in grayed box). Trace showing the execution latency of attacker's inference branch.}
	\label{fig:side_channel}
\end{figure}

\vspace{1mm}
\noindent\textbf{Evaluation on PoC of \mytitle side channel.}
\textcolor{black}{
We first evaluate the \mytitle side channel on a Proof-of-concept (PoC) attack. Listing~\ref{lst:overview} shows the major sources of the attacker and victim process for \mytitle-v1. The victim program maintains a secretive and a non-secretive array (e.g., {$sec$} and {$array1$} respectively). It also contains a trigger gadget ($b_{v0}$) and corresponding transmitter gadget (i.e., conditional branch, $b_v$). If the parent branch of $b_v$ is mis-speculated, $b_v$ could be executed and resolved speculatively. In the speculative domain, out-of-bound access from $array1$ could lead to load of an element from the secretive data structure $sec$, which alters the PHT entry state for $b_v$.
The attacker can infer the value of one element once in $sec$ through observing the speculative update of the targeted PHT entry. Note {$preset\_ghr()$} is required only when using history-based prediction. The attacker can then exfiltrate the content of the entire array by changing the index value to $array1$.
Figure~\ref{fig:side_channel} shows the observed latency of $b_a$ execution for multiple iterations of the attack under one-level prediction (Figure~\ref{fig:side_channel_one}) and history-based prediction (Figure~\ref{fig:side_channel_two}). 
Specifically, the execution latency of $b_a$ during the $2^{nd}$ inference (for 1-level predictor) or the $4^{th}$ inference (for history-based predictor) directly correlates with the outcome of $b_v$ and hence, the value of the corresponding element in the $sec$ array that was speculatively loaded. The results demonstrate that the attacker can successfully recover the exact values of the secretive array (i.e., $N$ represents value $1$ and $T$ represents $0$) \emph{even when there is no explicit control flow or data flow dependent on it}. 
}
\textcolor{black}{
Table~\ref{table:performance} lists the transmission rate and accuracy of the attack using the two branch prediction modes. Both attacks can achieve very high raw bit accuracy. Also, as expected, the side channel attack under history-based prediction can achieve much higher bit rate compared to one-level prediction (188Kbps vs. 0.3kbps) as prediction model re-enforcement is not needed in the history-based prediction mode. }
\begin{table}[t]
    \setlength\tabcolsep{3.5pt}
    \centering
	\footnotesize
    \begin{tabular}{lcc}
    \toprule
    \textbf{Predictor mode}    & \textbf{Transmission rate} ($Kbps$)   & \textbf{Accuracy} ($\%$)\\
    \midrule
    \textbf{One-level} &   0.3   &   99.8 \\
    \textbf{History-based} &  188 & 98.4\\
    \bottomrule
    \end{tabular}
    \caption{\mytitle side channel performance.}
    \label{table:performance}
\end{table}

\textcolor{black}{
Finally, it is important to note that the code gadget exploited by \mytitle is not vulnerable to cache-based transient execution attacks (e.g., Spectre V1). This is because there is no distinguishable data dependency on the sensitive data $sec$ as $sum$ is accessed in both branch directions. Additionally, even though there is a control-flow dependency on the speculative-loaded memory ($if$ in line \textcolor{black}{11} and $else$ in line \textcolor{black}{13}), the corresponding two basic blocks are mapped to the same instruction memory line, making it impossible to be inferred using cache attack such as Flush+Reload~\cite{flush_reload,kocher2019spectre}. In contrast, \mytitle is able to infer secret with this gadget by directly observing the speculative update of $b_v$'s PHT entry, regardless of instruction memory layout and data access patterns.}
Note that a \mytitle-v2 variant can be implemented similarly, with the only change of replacing the trigger gadget ($b_{v0}$) with an indirect jump leading to the transmitter gadget.

\subsection{\mytitle Code Gadget Analysis}
\label{subsec:real_world_impl}

\mytitle exhibits several main characteristics that make it different from the classical Spectre attacks. Firstly, the proposed attack completely relies on the BPU for both accessing and inferring secrets in speculative execution path. This attack can further extend the attack surface of previously demonstrated transient execution attacks with the majority of which relying on exploiting the cache hierarchy as the secret transmitting medium~\cite{xiong2020survey}. Secondly, 
the code pattern in \mytitle can utilize code gadgets potentially widely existing in normal code base. For example, Spectre V1 uses \emph{memory indirection} where an out-of-bound accessed value is used as the index to access another memory location. However, such pattern has been rarely found even in large-scale benign code bases~\cite{google-zero}. Notably, the new attack variant exploits a simpler transmitter gadget (e.g., a branch execution based on speculation-derived conditionals), which makes it easier to find exploitable gadgets in the victim’s code base. 

\begin{table}[t]
\small
\centering
\begin{tabular}{|p{0.36\textwidth}|}
\hline
\begin{lstlisting}[frame=none,aboveskip=-2pt,belowskip=-3pt]
if (x < bound) return; // |\textcolor{darkgreen}{\textbf{$b_{v0}$}}|
if (array1[x]) // |\textcolor{darkgreen}{\textbf{$b_{v}$}}|
    <some_operations>;
\end{lstlisting}
\begin{center}(a) Consecutive conditional branches\end{center}
\\
[-1.5ex]\hline\\[-2ex]
\begin{lstlisting}[frame=none,aboveskip=-2pt,belowskip=-3pt]
if (x < bound) // |\textcolor{darkgreen}{\textbf{$b_{v0}$}}|
    for (int i = 0; i < bound; i++)
        if (array1[x + i]) // |\textcolor{darkgreen}{\textbf{$b_{v}$}}|
            <some_operations>;
\end{lstlisting}
\begin{center}(b) Multi-level speculation\end{center}
\\
[-1.5ex]\hline\\[-2ex]
\begin{lstlisting}[frame=none,aboveskip=-2pt,belowskip=-3pt]
for (int i = x; i < bound; i++) // |\textcolor{darkgreen}{\textbf{$b_{v0}$}}|
    if (array1[i])  // |\textcolor{darkgreen}{\textbf{$b_{v}$}}|
        <some_operations>;
\end{lstlisting}
\begin{center}(c) Loop-based speculation\end{center}
\\\hline
\end{tabular}

\begin{lstlisting}[caption={Example patterns vulnerable to \mytitle-v1.},frame=none,label={lst:side_channel}]
\end{lstlisting}
\end{table}

\begin{figure}[t]
	\centering
	\includegraphics[width=0.37\textwidth]{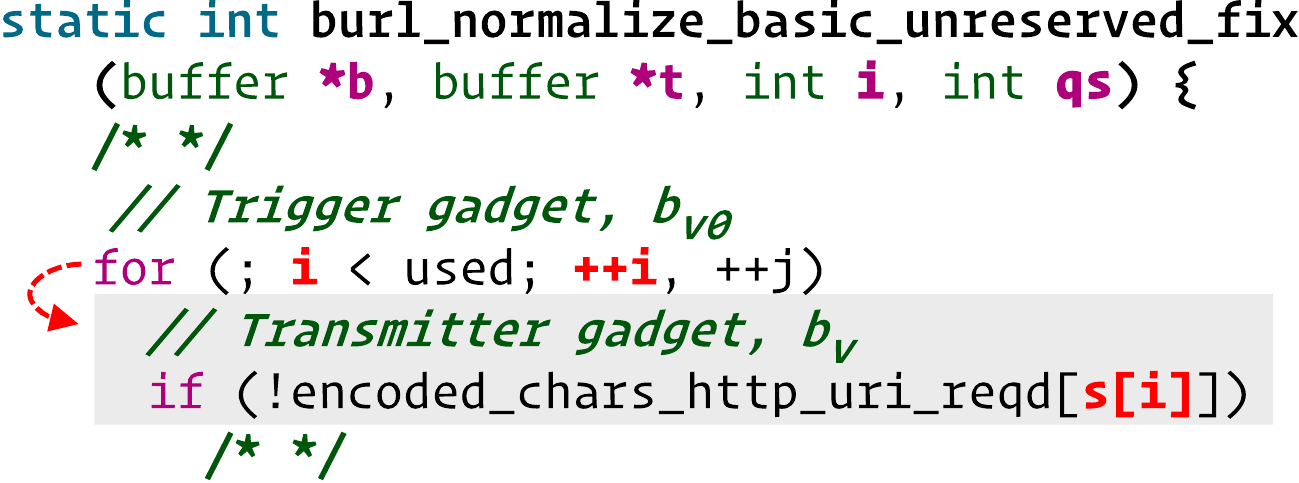}
	\caption{Potential \mytitle-v1 code gadget found in \textcolor{black}{\texttt{burl.c}} of \textbf{lighttpd}.}
	\label{fig:spectre_v1_gadget}
\end{figure}

\vspace{1mm}
\noindent\textbf{\mytitle-v1 gadget in real-world applications.} The key pattern of the \mytitle-v1 attack is nested speculation of a conditional branch where its earlier branch has been mis-speculated. 
Listing~\ref{lst:side_channel} illustrates several examples of vulnerable code patterns that can be potentially leveraged. In each of the patterns, the first branch is used to trigger the transient execution of the target branch. Note that the expression itself used as the condition of the nested branch may not be confidential in the victim’s program semantic. But speculation can potentially enable access to unintended data through it (e.g., out-of-bound access). This makes \mytitle-v1 even more dangerous than prior branch predictor based exploitation that requires secret-dependent branching in victim’s binary. 
We perform static program analysis to find potential \mytitle-v1 gadgets in real-world applications. 
Figure~\ref{fig:spectre_v1_gadget} shows a potential \mytitle-v1 in \textcolor{black}{the \texttt{burl} module of {lighttpd}~\cite{lighttpd} web server. This function is used to parse the URL path of an HTTP request.} In this function, the $if$ branch in line \textcolor{black}{$68$} can be speculatively executed with the function argument $i$ value that is out of the boundary of a \textcolor{black}{char} array $s[.]$.
As a result, partial information about the data in speculative domain can be inferred through the PHT entry state that is altered. We note that an exhaustive search of \mytitle-v1 gadget requires comprehensive tainting and dynamic program analysis (such as symbolic execution)~\cite{fastspec,kleespectre}, which is out of the scope of this work.

\begin{table}[t]
    \setlength\tabcolsep{2.5pt}
    \centering
    \footnotesize
    \begin{tabular}{lcc||lcc}
    \toprule
    \textbf{Binary} &  \textbf{\mytitle-v2} & \textbf{SS} & \textbf{Binary} &  \textbf{\mytitle-v2} & \textbf{SS}\\
    \midrule
    {apache2} & 6393 & 1639  &    {glibc} & 8235 & 3422 \\
    {ld} & 375 & 138   &    {libcrypto} &  2872 & 2295  \\
    {libssl} & 5664 & 1408   &    {lighttpd} & 5474 & 3984  \\
    {nginx} & 6279 & 2185  &    {openssh} & 7674  &  408 \\
    {pthread} &  218 & 64 &    {stdc} & 1572 & 720 \\
    \bottomrule
    \end{tabular}
    \caption{Number of \mytitle-v2 transmitter gadgets using memory load from any of the callee-save registers. \emph{SS} denotes SmotherSpectre.}
    \label{table:gadget}
\end{table}

\vspace{1mm}
\noindent\textbf{\mytitle-v2 gadget analysis.} \mytitle-v2 can be more dangerous than \mytitle-v1 due to the possibility of chaining arbitrary code sequence. Spectre V2 implementations typically use cache side channel in the transmitter gadget. 
Note that \mytitle-v2 is considerably less stringent for the exploitability of the gadgets: \emph{it only requires one conditional branch as the transmitter gadget}.
To understand the advantage of \mytitle-v2 compared to other attacks in terms of gadget availability, we analyze several common real-world applications and libraries to quantify the number of exploitable gadgets. We choose SmotherSpectre~\cite{SMoTherSpectre} as the baseline, which is the prominent non-cache Spectre V2 attack that exploits port contention patterns in the transmitter gadget. Same as~\cite{SMoTherSpectre}, we assume the targeted secrets are arguments of a function call in System V calling convention, which are stored in \texttt{RDI, RSI, RCX} and \texttt{RDX} in x86. As a result, a virtual function call (implemented as an indirect call) can be regarded as the trigger gadget. 
We perform extensive search on the transmitter gadget in common software binaries ranging from server applications to widely used C libraries. For both attacks, the transmitter gadget needs to start with a branch using a memory value pointed by any of these registers (or registers tainted by them) as the conditional. In this experiment, \textcolor{black}{we only consider \texttt{TEST}$\rightarrow$\texttt{JXX} sequence that allows leakage of information for branch conditional at the bit level.} Different from \mytitle-v2, the transmitter gadget in SmotherSpectre (\textit{SS}) has to exhibit distinctive port contentions between the taken and fall-through path of the target branch $b_v$. 
Table~\ref{table:gadget} shows the total number of transmitter gadgets found for \mytitle-v2 and SmotherSpectre. These gadgets either leak the value of the aforementioned registers or the value of memory location pointed by them. As we can see, by exploiting speculative updates in PHT, the number of code gadgets exploitable by \mytitle-v2 is in total $2.75$ times the one for SmotherSpectre. 

\begin{figure}[t]
	\centering
	\includegraphics[width=0.485\textwidth]{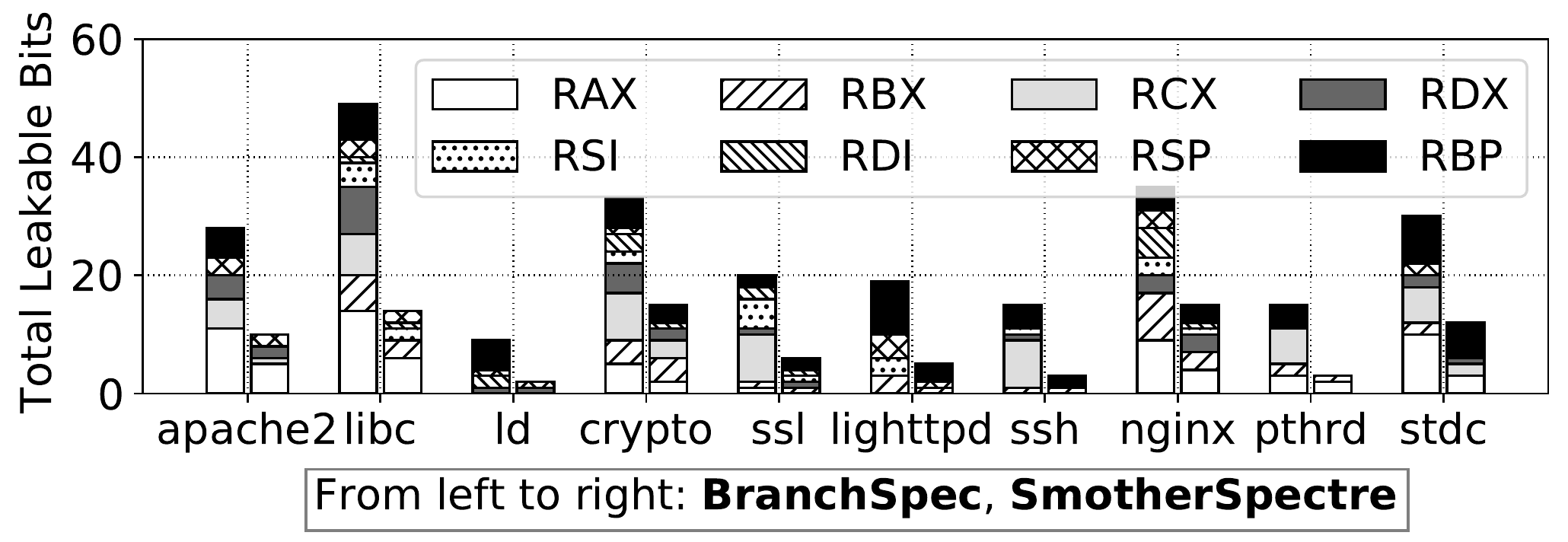}
	\caption{Breakdown of leakable bit locations through different registers for \mytitle-v2 and SmotherSpectre.}
	\label{fig:leakable_bit}
\end{figure}

Since conditional branches with 
\texttt{TEST} typically only compares certain bits of an operand, we further investigate the locations of different bits within an 8-byte memory value that could be leaked. 
Figure~\ref{fig:leakable_bit} shows the number of leakable bit offsets our attack can achieve compared to SmotherSpectre via each specific register. 
It is observed that \mytitle leaks significantly more bits compared to SmotherSpectre. 
\textcolor{black}{Overall, \mytitle can leak $2\times$ more bits compared to SmotherSpectre across the analyzed binaries/libraries.}
Note that neither attack is dependent on cache timing channel, making them resistant against most of the proposed defense mechanisms against cache timing channel.

\begin{figure}[t]
	\centering
    \subfloat[b][\texttt{do\_cipher} virtual function call]{
	\includegraphics[width=0.45\textwidth]{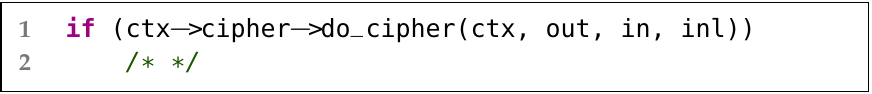}
	\label{fig:v2_c}}

	\subfloat[b][Assembly representation of Listing~\ref{fig:v2_gadget}a]{
	\includegraphics[width=0.45\textwidth]{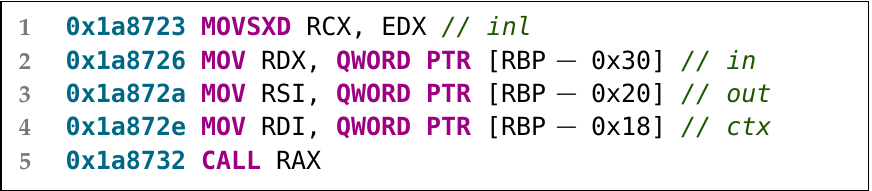}
	\label{fig:v2_asm}}
	 \captionsetup{type=lstlisting}
	\caption{\mytitle-v2 triggering gadget from OpenSSL in \texttt{evp\_EncryptDecryptUpdate} function.}
	\label{fig:v2_gadget}
\end{figure}

\begin{figure}[t]
	\centering
    \subfloat[b][Leaking the 2nd bit]{
	\includegraphics[width=0.24\textwidth]{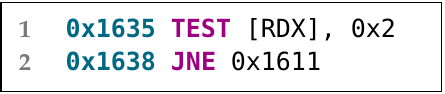}
	\label{lst:v2_gadget1}}
	\subfloat[b][Leaking the 6th bit]{
	\includegraphics[width=0.24\textwidth]{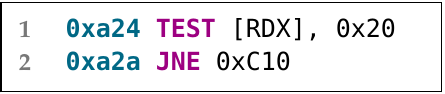}
	\label{lst:v2_gadget2}}
	\captionsetup{type=lstlisting}
	\caption{\mytitle-v2 transmitter gadgets in OpenSSL leaking certain bit of the first byte of secret pointed by \texttt{RDX}.}
	\label{lst:transmitter}
\end{figure}

\subsection{Attack Case Study on OpenSSL}
\label{subsec:real_atk}
As a demonstration, we show how an \textcolor{black}{unprivileged} attacker can leak the secretive plaintext from a victim process running encryption using the OpenSSL. We assume the attacker process runs on the same physical core as the victim process (using SMT context).
Listing~\ref{fig:v2_gadget}a shows the triggering gadget used for \mytitle-v2 with the assembly representation showing in Listing~\ref{fig:v2_gadget}b. The code gadget is within the implementation of the \texttt{EVP\_EncryptDecryptUpdate} function of OpenSSL. 
Specifically, we target the \texttt{do\_cipher} virtual function call. Note that the same vulnerability is exploited via port contention in~\cite{SMoTherSpectre}.
The secret (i.e., a victim's plaintext in this case) is passed as a pointer using the $3^{rd}$ argument ($in$) to the function via \texttt{RDX} register.
\textcolor{black}{We observe that enough deterministic $taken$ branches can be collected by the attacker to prepare the GHR for PHT collision.}
The attack steps are as follows:
1) the attacker performs branch target injection~\cite{kocher2019spectre} from its own address space to poison the indirect call in the triggering gadget to divert the control flow to the transmitter gadget. Meanwhile, the attacker also performs the target PHT initialization for history-based predictor mode, as discussed in Section~\ref{ss:side_channl_impl}. 2) The attacker triggers the encryption in the victim application, which eventually invokes \texttt{EVP\_EncryptDecryptUpdate}. When the triggering gadget is encountered, the speculative control flow transfers to the transmitter gadget that alters the pre-initialized PHT entry based on the secret data. Note that here \texttt{RDX} points to the memory location of the secret in the speculative domain. 3) The attacker recovers the secret by inferring the target PHT entry similar to step 3 in Section~\ref{ss:side_channl_impl}. 

Listing~\ref{lst:transmitter} illustrates two example transmitter gadgets found in the OpenSSL library. \textcolor{black}{The conditional branch in each of the transmitter gadgets will resolve as \emph{Taken} if the leaking bit---the second (Listing~\ref{lst:transmitter}a) or sixth (Listing~\ref{lst:transmitter}b) \emph{least significant bit} of the $1^{st}$ secret byte---is `0'. The same branch will be resolved as \emph{Not Taken} speculatively if the corresponding secretive bit is `1'.} Note that more bits of the secret could be exfiltrated by utilizing other transmitter gadgets that leak bits at different locations. 
\textcolor{black}{
From our analysis, we have found that using the same triggering gadget, the plausible transmitter gadgets in OpenSSL can collectively leak all 8 bits of the first byte of the secret. Therefore, by chaining an additional control gadget (e.g., through ROP-like chaining~\cite{bhattacharyya2020specrop}) that gradually shifts the offset to the address of the secret, all bytes in the secret can be leaked. 
We run the attack with the identified transmitter gadgets 1000 times, and our results show that the attack can successfully infer individual bits of the secrets with an average accuracy of 97.3\%.
}

\subsection{\mytitle in the Context of SGX}
While we mainly consider a conventional threat model where attackers only have userspace accesses, \mytitle can be easily extended to trusted execution environments (e.g., Intel SGX) with the assumption of a privileged attacker (i.e., control over the OS). 
Under such scenario, the attacker can perform fine-grained scheduling of the victim's execution. We note that \mytitle can be more effective in the context of SGX for the following reasons: First, the high-resolution stepping (i.e., through interrupts) allows the attacker to probe the victim's speculative PHT updates \emph{soon} after the transient execution of the victim's targeted branch, effectively denoising the inference operation. It is worth noting that a precise interrupt right at the target branch is not possible as it is in the transient execution path. Second, the privileged attacker can replay a faulting instruction (e.g., page-fault a load~\cite{skarlatos2019microscope}) before the victim's triggering gadget. This will enable \mytitle-v2 to integrate different transmitter gadgets over multiple rounds and potentially leak all bits of secrets in one run of the victim.

\section{Discussions}

\subsection{Effectiveness of Existing Countermeasures}

\vspace{1mm}
\noindent\textbf{System-level defenses.} 
\textcolor{black}{The industry has rolled out several mitigation patches to limit transient execution attacks. Notably, microcode updates can prevent BTB-poisoning to higher-privilege code (IBRS~\cite{ibrs}), eliminate indirect branch predictor sharing across SMT threads (STIBP~\cite{stibp}), and flush BTB upon context switches (IBPB~\cite{ibpb}). Retpoline replaces indirect jumps with return trampoline to trap speculation~\cite{turner2018retpoline,retpoline_intel}. Similar to classical Spectre V2, \mytitle-v2 may be mitigated with the use of STIBP and IBPB since they restrict cross-process branch target poisoning. However, software vendors have revealed that these mechanisms essentially disable indirect branch prediction entirely~\cite{seccomp} and are reluctant to enable them widely~\cite{Taming}. In fact, recent works find that such defenses have rarely been adopted in user-level applications~\cite{SMoTherSpectre,evtyushkin2018branchscope}, leaving the \emph{same-process} or \emph{cross-process} userspace attack still viable.
}
\textcolor{black}{
Furthermore, while avoiding secret dependent branching (as practiced in crypto algorithm implementations~\cite{gnupg,chowdhuryyseeds}) can mitigate BPU attacks in non-speculative domain, such mechanism does not protect \mytitle as other irrelevant branches could be potentially exploited to access the secret speculatively and use it as the branch predicate (which defies program semantic). Finally, fencing as a general speculation protection technique can mitigate propagation of speculative secrets to the PHT. However, 
adding fences to all branch instructions can incur significant performance penalty~\cite{phoronix}.
}

\vspace{1mm}
\noindent\textcolor{black}{\textbf{Architecture-level defenses.} Researchers are continuously proposing architecture-level defense mechanisms to mitigate the transient execution attacks. As the majority of these attacks primarily rely on cache to leak secret information, thwarting the cache side channel is the main focus of many architecture-level defense proposals~\cite{invispec,saileshwar2019cleanupspec,MuonTrap,khasawneh2019safespec}. 
Since the proposed attack does not rely on the existence of cache side channel, these defenses are ineffective against \mytitle. 
\textcolor{black}{Oblivious speculation frameworks prevent speculatively accessed data by restricting value propagation to microarchitecture states~\cite{stt,sda,weisse2019nda,li2019conditional}. These defenses ensure that microarchitecture state changes are only in place eventually for legitimate instruction executions. These techniques can defend against \mytitle if they are applied to audit microarchitecture states in PHT.} 
\textcolor{black}{
Similar to secure cache mechanisms, randomization and isolation-based schemes for branch predictors have been proposed to eliminate cross-process BPU side channels~\cite{vougioukas2019brb,lee2020securing,zhao2020lightweight}. 
These mechanisms in general limit branch predictors as the transmitter medium for information leakage, and thus can potentially mitigate the proposed attack. However, such designs typically involve considerable performance overhead and non-trivial hardware cost. 
}
}

\subsection{Potential Future Mitigations}
We now discuss several viable mitigation techniques that can be employed in future systems.

\vspace{1mm}
\noindent\textbf{Delaying PHT update in speculation.} The PHT update for a branch instruction resolution can be delayed until the instruction is committed or can no longer be squashed by any other prior instructions. This way, the PHT will not be impacted by the branches that are resolved in the wrong path of speculation, which avoids the possibility that unintended program paths influence the state of the PHT. However, prior studies have shown early BPU updates (at resolution time) exhibit performance benefits compared to commit-time update~\cite{yale}. Moreover, updates with respect to both wrong-path and correct-path branch resolutions have such positive effect. The underlying reason is that resolve-time update of a branch brings \textit{ahead-of-time} update of PHT, which can enable fast correction of wrong prediction decisions for branch instructions in-flight. In fact, we experimentally found that branch predictors with resolve-time updates of the PHT brings on average 8\% performance gain (for SPEC benchmarks) compared to resolve-time updates. Given that modern branch predictors are highly optimized hardware components, such performance difference is non-trivial. Thus, it is necessary to explore defensive strategies that retrain the performance benefit of speculatively updated processors while also ensuring information security.

\vspace{1mm}
\noindent\textbf{PHT states obfuscation for transient branches.} 
One straightforward mitigation against this attack is to rollback the state changes in pattern history once a mis-speculation is handled. This can be achieved by initiating checkpointing mechanisms to record the state of PHT before speculation. To do so, one checkpoint has to be created for each speculative branching point. Such mechanism can bring considerable cost of tracking the PHT state changes. This is especially the case for modern deep-pipelined processors where multi-level nested speculation is common. 
One potential solution to alleviate the restoration overhead is to mark the PHT entries that have been altered by transient branch instructions. However, instead of restoring those PHT entries to the original states, we could choose to obfuscate those PHT entries once speculation is terminated. Such technique is motivated by prior studies showing that perfectly checkpointing the PHT for speculative updates does not bring noticeable performance benefits~\cite{yale}.

\vspace{1mm}
\noindent\textbf{Invisible PHT entry.} The processor can isolate a region of PHT for the speculative branches where it records and updates the PHT state speculatively. In this scheme, when a speculatively executed branch is resolved, the shadow-PHT will be updated instead of directly updating the PHT. 
This shadow-PHT will be visible to other speculative branches, thus retaining the benefits of speculative update of BPU states. 
Each entry in the shadow-PHT can be associated with a process id, and only the branches matching the process id will be able to use this shadow-PHT entry. This will also block same-process speculative information leakage. The size of this shadow-PHT can be minimal as it only needs to keep the PHT entries corresponding to branch instructions that are in ROB and resolved speculatively. When the branch corresponding to a shadow-PHT entry is committed, the PHT entry will be merged with the original entry. Note that conflict handling mechanism has to be implemented here since it may be possible that the same PHT entry is updated by two different processes, hence there are two copies of it in the shadow-PHT.

\section{Related Work}

Prior works have shown attacks exploiting BPUs to infer or transmit secrets in the \emph{non-speculative domain}~\cite{evtyushkin2016understanding,evtyushkin2018branchscope,lee2017inferring,huo2020bluethunder}. Specifically, Evtyushkin \textit{et al.}~\cite{evtyushkin2016understanding} propose a PHT-based covert channel that manipulates a large set of PHT entries to modulate timings of branch instruction executions. BranchScope~\cite{evtyushkin2018branchscope} shows a side channel that steals intended program secrets through observing PHT state changes. BlueThunder~\cite{huo2020bluethunder} extends the exploitation of one-level prediction in \cite{evtyushkin2018branchscope} to two-level prediction mode with fine-grained stepping of the victim process. Furthermore, the work in~\cite{mambretti2019two} has revealed that changes to the BTB during speculation are persisted. With such observation, it demonstrates a spectre attack variant that uses caches as the side channel transmitter to sense BTB changes and infer secrets.  
We note that \mytitle is substantially different as none of these works has investigated the security vulnerability of \emph{speculative PHT update} due to branch execution in speculative domain. 
\mytitle can exfiltrate unintended sensitive data by observing PHT state alterations due to branch resolutions in the transient execution path. With \mytitle, the transmitter gadget only needs a speculative execution of conditional branches. The end-to-end attack can be carried out by \emph{exploiting the BPU alone} for both triggering mis-speculation and transmitting speculative secrets. 
Our work further broadens the transient execution attack surface and motivates the need for rethinking the security of branch predictor design for speculative processors.

Transient execution attacks can be broadly categorized to \textit{Spectre}- and \textit{Meltdown}-type where the former relies on speculation due to prediction and the latter depends on speculation due to execution handling during retirement~\cite{xiong2020survey}. \textcolor{black}{Previous \textit{Spectre}-type attacks use the BPU to trigger the speculation~\cite{kocher2019spectre,canella2019systematic}, while other hardware-based side channel exploits are responsible for the actual data transmission (e.g.., using caches~\cite{kocher2019spectre} and port contentions~\cite{SMoTherSpectre}).
Many research works have proposed techniques to eradicate speculation footprint on caches as countermeasures while allowing speculation to continue safely~\cite{invispec,saileshwar2019cleanupspec,MuonTrap,khasawneh2019safespec}. Different from these hardware-specific defenses, oblivious speculation frameworks~\cite{weisse2019nda,stt,li2019conditional,khasawneh2019safespec,barber2019specshield,value_pred} propose unified techniques for defeating transient execution attacks by tracking the propagation of speculatively accessed secret and delaying microarchitecture state changes by any instructions until they are destined to be committed. There are done at the expense of incurring various levels of performance degradation.} While these generic secure speculation frameworks can be applied to mitigate \mytitle attacks, our preliminary studies show that the branch predictor's performance can be very sensitive to the timing of its internal state update. Therefore, there is potentially a need to design secure mechanisms that can retain the performance advantage of speculative BPU state updates.

\section{Conclusion}
In this work, we present \mytitle, an attack framework that exploits branch predictors as the transmitting medium in speculative domain in modern processors. Our key finding is that transient executions of conditional branches introduce alterations in the PHT, which will remain after these branch instructions are squashed. This allows an attacker to infer unintended secrets if speculatively executed conditional branches depend on them. Leveraging this vulnerability, we demonstrate a high bit rate speculative covert channel attack as well as V1/V2-type \mytitle side channel attacks. The new side channels entirely rely on exploitation of branch predictor unit and can take advantage of code patterns much simpler than the ones used in Spectre attacks. We perform code gadget analysis on several popular software code bases and demonstrate a real-world attack on OpenSSL. Our evaluation results show that \mytitle exhibits higher attack capability than existing works. Finally, we discuss potential mitigation mechanisms that secure branch predictors against transient execution attacks.  

\section*{Acknowledgments}
This material is based upon work supported in part by U.S. National Science Foundation under CNS-2008339.

\ifCLASSOPTIONcaptionsoff
  \newpage
\fi

\bibliographystyle{IEEEtran}
\bibliography{main}

\end{document}